\documentclass[twocolumn,aps,prb,10pt,superscriptaddress]{revtex4-2}
\usepackage{graphicx}
\usepackage{dcolumn}
\usepackage{bm}
\usepackage{multirow}

\usepackage{amssymb}
\usepackage{amsmath}
\usepackage{graphicx}
\usepackage{xcolor}
\usepackage{braket}
\usepackage{CJK}
\usepackage{indentfirst}
\usepackage{cases}
\usepackage{booktabs}
\usepackage{tabularx}
\usepackage{enumitem}

\def\Eq#1{Eq.~(\ref{#1})}

\def\avg#1{\left\langle#1\right\rangle}

\def \kuohao#1{\left(#1\right)}
\def \fkuohao#1{\left[#1\right]}

\def \abs#1{\left|#1\right|}

\newcommand{\expup}[1]{\mathrm{e}^{#1}}

\def \mat#1{\begin{pmatrix}#1\end{pmatrix}}

\newcommand{\nota}[1]{\texorpdfstring{$#1$}{}}
\newcommand{\equa}[1]{\texorpdfstring{
    \begin{equation}
        #1
    \end{equation}
}{}}

\newcommand{\bea}{\begin{eqnarray}}
\newcommand{\eea}{\end{eqnarray}}

\definecolor{UkiyoRed}{RGB}{223,126,102}
\definecolor{UkiyoGreen}{RGB}{148,181,148}
\definecolor{UkiyoYellow}{RGB}{237,199,117}
\definecolor{YinshuaiBlue}{RGB}{0,125,255}
\definecolor{YinshuaiPink}{RGB}{255,111,180}
\definecolor{YinshuaiYellow}{RGB}{255,177,0}
\usepackage[
    colorlinks,
    citecolor=YinshuaiPink,
    linkcolor=YinshuaiBlue,
    urlcolor=YinshuaiYellow,
]{hyperref}

\newcommand{\subref}[1]{\textcolor{YinshuaiBlue}{#1}}

\usepackage{slashed}
\usepackage{hhline}

\begin{document}
\title{Preempting Fermion Sign Problem: Unveiling Quantum Criticality through Nonequilibrium Dynamics in Imaginary Time}

\author{Yin-Kai Yu}
\affiliation{School of Physics, Sun Yat-sen University, Guangzhou 510275, China}
\affiliation{Guangdong Provincial Key Laboratory of Magnetoelectric Physics and Devices, Sun Yat-sen University, Guangzhou 510275, China}
\affiliation{Beijing National Laboratory for Condensed Matter Physics \& Institute of Physics, Chinese Academy of Sciences, Beijing 100190, China}
\affiliation{University of Chinese Academy of Sciences, Beijing 100049, China}

\author{Zhi-Xuan Li}
\affiliation{School of Physics, Sun Yat-sen University, Guangzhou 510275, China}
\affiliation{Guangdong Provincial Key Laboratory of Magnetoelectric Physics and Devices, Sun Yat-sen University, Guangzhou 510275, China}

\author{Shuai Yin}
\email{yinsh6@mail.sysu.edu.cn}
\affiliation{School of Physics, Sun Yat-sen University, Guangzhou 510275, China}
\affiliation{Guangdong Provincial Key Laboratory of Magnetoelectric Physics and Devices, Sun Yat-sen University, Guangzhou 510275, China}

\author{Zi-Xiang Li}
\email{zixiangli@iphy.ac.cn}
\affiliation{Beijing National Laboratory for Condensed Matter Physics \& Institute of Physics, Chinese Academy of Sciences, Beijing 100190, China}
\affiliation{University of Chinese Academy of Sciences, Beijing 100049, China}

\date{\today}
\begin{abstract}
The notorious fermion sign problem, arising from fermion statistics, presents a fundamental obstacle to the numerical simulation of quantum many-body systems. Here, we introduce a framework that circumvents the sign problem in the studies of quantum criticality and its associated phases by leveraging imaginary-time nonequilibrium critical dynamics. We demonstrate that the critical properties can be accurately determined from the system’s short-time relaxation, a regime where the sign problem remains manageable for quantum Monte-Carlo (QMC) simulations. After validating this approach on two benchmark fermionic models, we apply it to the sign-problematic Hubbard model hosting SU(3)-symmetric Dirac fermions. We present the first numerically exact characterization of its quantum phase diagram, revealing a continuous transition between a Dirac semi-metal and a SU(3) antiferromagnetic phase. This transition defines an unconventional Gross-Neveu universality class that fundamentally reshapes current understanding of Gross-Neveu criticality. Our work provides a powerful tool for investigating sign-problematic systems and quantum criticality.

\end{abstract}
\maketitle

\pdfbookmark[1]{Introduction}{sec1}
\noindent\textcolor{YinshuaiBlue}{\textbf{INTRODUCTION}}\\
The complex sign structure of the quantum mechanical wavefunctions presents a central dichotomy in many-body physics. On the one hand, it endows fertile exotic phenomena across condensed matter~\cite{Sachdevbook,zaanen2008science} and high energy physics~\cite{weinbergBook}. Quantum Monte-Carlo (QMC) is among the most important theoretical approaches to study the strongly correlated quantum many-body systems~\cite{BSS,Hirsch1981PRL,Hirsch1985PRB}, offering numerically exact and unbiased solutions. On the other hand, the sign structure results in the notorious sign problem~\cite{Loh1990PRB,Sandvik2000PRB,Troyer2005sign,Hastings2016,Ringel2017SA}, which considerably plagues the application of QMC to investigating interacting models potentially featuring intriguing physics, including the Hubbard model at generic filling~\cite{Hubbard1,Hubbard2} and the lattice QCD at finite baryon density~\cite{BooklatticeQCD}. Since the general solution of the sign problem is lacking and nondeterministic polynomial (NP)-hard~\cite{Troyer2005sign}, developing generic strategies to circumvent or mitigate it will definitely provoke significant advances in the realm of quantum many-body physics~\cite{li2019review,wiese1999prl,huffman2014prb,ZhangSC2005TRS,Sachdev2012Science,li_solving_2015,li_majorana-time-reversal_2016,Zhang2003prl,wang2015prl,xiang2016prl,mondaini_quantum_2022,Hangleiter2020SA,Ryan2021PRL,Han2024arXiv,Alexandru2022RMP,Ringel2020PRR,Chang2024PRB,Yao2022prb,Meng2022PRB,Assaad2021PRB,Abolhassan2021PRL,Berg2023PRL}.

Among various exotic phenomena in quantum many-body systems,  quantum criticality emerges as a particularly crucial and fascinating one.  The underlying physics of the quantum critical point (QCP) lays the foundation for achieving a unified theoretical framework to characterize different phases and phase transitions~\cite{Sondhi1997RMP,Vojta2003Review,Hertz1976prb,Millis1993prb,Berg2019QCPReview}. Moreover, quantum criticality is intimately associated with fundamentally important phenomena in condensed matter physics such as high-$T_c$ superconductivity~\cite{Scalapino2012RMP,Wen2006rmp,Kivelson2015PRL} and strange metallicity~\cite{Varma2020rmp,PhilipReviewScience}. The understanding of the QCP in a non-perturbative way is severely hindered by the sign problem likewise.  

\begin{figure*}[t]
    \centering
    \includegraphics[width=\linewidth]{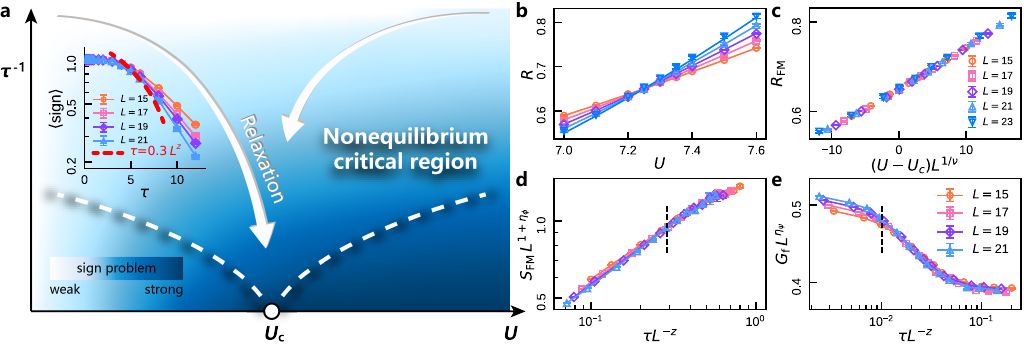}
    \caption{\textbf{Scheme of preempting sign problem to probe quantum criticality via short-time critical dynamics and the application in single-Dirac-fermion Hubbard model.} 
    \textbf{a}, With some typical initial states, scaling behaviors governed by the QCP are reflected in the short-time stage, in which the sign problem is still weak as shown in the inset. The white dashed line indicates the nonequilibrium critical region associated with the QCP. The red dashed line in the inset indicates the average sign at $\tau=0.3L^z$, where $z=1$ due to the nature of the Dirac QCP. 
    \textbf{b}, Determination of QCP as $U_c = 7.220(37)$ via the intersection points of curves of the correlation length ratio $R_{\rm FM}$ versus $U$ for different sizes at $\tau=0.3L^z$ with DSM initial state. 
    \textbf{c}, Determination of $1/\nu = 1.18(3)$ by scaling collapse of $R_{\rm FM}$ versus rescaled $(U - U_c)$, with a reduced chi-squared value $\chi_{\nu}^2=1.824$ indicating good quality.
    \textbf{d}, Determination of $\eta_\phi = 0.33(2)$ via scaling collapse of curves of the structure factor $S_{\rm FM}$ versus rescaled $\tau$ at $U_c$, with $\chi_{\nu}^2=0.606$. 
    \textbf{e}, Determination of $\eta_\psi = 0.135(2)$ via scaling collapse of curves of the fermion correlation $G_{\rm f}$ versus rescaled $\tau$ at $U_c$, with $\chi_{\nu}^2=1.241$. 
    The dashed lines in \textbf{d} and \textbf{e} indicate the boundary of the nonequilibrium scaling region; only the regions to the right are included in the scaling collapse analysis (see SM Sec.~\ref{Sec:S2-B}).}
    \label{fig:SLAC}
\end{figure*}

Here, we propose a general framework to preempt the sign problem in QMC and unveil the quantum criticality and associated ground-state phase diagram. It is widely acknowledged that fathoming the nonequilibrium properties in quantum many-body systems is substantially more challenging than equilibrium ones. However, we leverage the nonequilibrium behavior and demonstrate that the short-time critical dynamics, which was first proposed in classical systems~\cite{Janssen1989} and used to detect the classical critical behaviors~\cite{Lizb1995prl} and then generalized to imaginary-time dynamics of QCP~\cite{Yins2014prb}, provides an ingenious strategy to reliably probe quantum criticality in the presence of the sign problem, as illustrated in Fig.~\ref{fig:SLAC}\subref{a}. The underlying mechanism is that in the short-time stage of imaginary-time relaxation for some typical initial states, (1) universal scaling behaviors manifesting the quantum criticality appear~\cite{Yins2014prb,yu2023dirac}; (2) the sign problem remains mild in this stage, such that the reliable results are accessible by QMC simulation. Hence, bestowing the scaling of short-time dynamics~\cite{Janssen1989,Lizb1995prl,Yins2014prb,yu2023dirac}, we can accurately determine the location and  critical properties of the QCP, while the sign problem therein, whose severity generally increases exponentially with the imaginary time~\cite{Troyer2005sign,AssaadReview,Scalettar2015prb}, is largely alleviated compared with the equilibrium one involving long imaginary-time evolution.

In the following, we demonstrate the framework by studying two typical interacting fermionic models featuring Dirac QCP. We unambiguously show that accurate critical properties can be accessed through short-imaginary-time dynamics, despite the presence of the sign problem. More remarkably, we adopt the approach to investigate a paradigmatic strongly interacting model hosting Dirac fermions with $\rm SU(3)$ symmetry,  which is sign-problematic in any known algorithm. For the first time, we establish the ground-state phase diagram of the model and reveal the critical properties of the quantum phase transition between the Dirac semi-metal (DSM) and a $\lambda_8$-antiferromagnetic (AFM) phase. 
Intriguingly, the QCP defines an unconventional Gross-Neveu transition that goes beyond the previous paradigms of O($N$) ordering, distinct from the known universality classes of Gross-Neveu transition.

\pdfbookmark[1]{Results}{Results}
\noindent\textcolor{YinshuaiBlue}{\textbf{RESULTS}}

\pdfbookmark[2]{Theoretical framework}{sec2}
\noindent {\it Theoretical framework} --- \\
We consider the imaginary-time relaxation dynamics for which the wave function $|\psi(\tau)\rangle$ evolves according to the imaginary-time Schr\"{o}dinger equation $-\frac{\partial}{\partial \tau} |\psi(\tau)\rangle=H|\psi(\tau)\rangle$
imposed by the normalization condition. As its long-time solution is the ground state, this equation forms the basis of the zero-temperature projector QMC (PQMC) method, which provides a routine approach to accessing ground-state properties in numerical computations~\cite{SM,Sorella1989EPL,Sorella1991IJMPB,AssaadReview}. Moreover, near the QCP, it was shown that universal scaling behaviors manifest themselves not only at the ground state, but also in the imaginary-time relaxation process. Particularly, with initial states corresponding to the fixed points of scale transformation, after a transient nonuniversal time scale, the short-imaginary-time dynamics of observable $O$ satisfies the following scaling form~\cite{Yins2014prb,yu2023dirac}:
\bea
O(\tau,g,L) = L^{-\kappa}f_O(gL^{\frac{1}{\nu}},\tau L^{-z}),
\label{scaling}
\eea
where $L$ is linear system size and $g$ is the distance to the critical point, $\kappa$ is the scaling dimension of $O$, and $\nu$ is the correlation-length exponent, $z$ is the dynamical exponent which is generally one in the Dirac QCP. Note that all critical exponents in Eq.~(\ref{scaling}) are controlled by the QCP in the ground state. Accordingly, from \Eq{scaling}, one can achieve the location of QCP and critical properties from the expectation value of $O$ in the relaxation process. This nonequilibrium approach is notably different from the conventional PQMC which requires that the ground state should be determined through the evolution of long imaginary time~\cite{SM,Sorella1989EPL,Sorella1991IJMPB,AssaadReview}. Because in general the severity of the sign problem exponentially increases with imaginary time in QMC~\cite{Troyer2005sign,Scalettar2015prb}, the sign problem in the short-time stage of relaxation is significantly mitigated compared with the direct simulation on ground-state properties, thereby enabling the large-scale QMC simulation with high accuracy despite the presence of the sign problem. In the following, we elucidate the theoretical framework by systematically studying several representative models with the sign problem. 

\pdfbookmark[2]{Single-Dirac-fermion Hubbard model}{sec3}
\noindent {\it Single-Dirac-fermion Hubbard model} --- \\
We first consider the SLAC fermion Hubbard model with the Hamiltonian~\cite{Vaezi2022prl,Li2019SA,Lang2019prl}:
\bea
H = \sum_{i,R} t_{R} c^{\dagger}_{i\uparrow} c_{i+R\downarrow} + \mathrm{h.c.} + U \sum_{i}\left(n_{i\uparrow}-\frac{1}{2}\right)\left(n_{i\downarrow}-\frac{1}{2}\right),\nonumber\\
\label{Ham1}
\eea
where $t_{R}= \frac{i(-1)^{R_x}}{\frac{L}{\pi}\sin{\pi\frac{R_x}{L}}} \delta_{R_y,0} + \frac{(-1)^{R_y}}{\frac{L}{\pi}\sin{\pi\frac{R_y}{L}}} \delta_{R_x,0}$ is the amplitude of long-range hopping on the square lattice and $U$ is the strength of repulsive interaction. 
The model is sign-problematic in any known QMC algorithm. Nonetheless, since the sign problem is relatively benign, a previous QMC study reveals a chiral Ising QCP separating the DSM phase and ferromagnetic (FM) phase~\cite{Vaezi2022prl}. 

\begin{figure}[t]
    \centering
    \includegraphics[width=\linewidth]{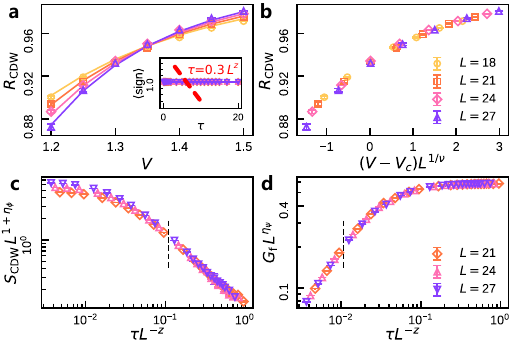}
    \caption{\textbf{Probing quantum criticality via short-imaginary-time critical dynamics in spinless $t$-$V$ model with CDW initial state.}  
    \textbf{a}, Determination of QCP as $V_c=1.35(1)$ via the intersection points of $R_{\rm CDW}$ versus $V$ for different $L$ at $\tau=0.3L^z$, where $z=1$. Shown in Inset is the evolution of average sign with red dashed curve marks $\tau=0.3L^z$. 
    \textbf{b}, Determination of $1/\nu=1.30(18)$ via scaling collapse of $R_{\rm CDW}$ versus rescaled $(V-V_c)$, with $\chi_{\nu}^2=1.518$ demonstrating the collapse quality.  
    \textbf{c}-\textbf{d}, Determination of $\eta_\phi = 0.49(5)$ and $\eta_\psi = 0.073(4)$ via scaling collapse of the short-imaginary-time dynamics of $S_{\rm CDW}$ and $G_{\rm f}$ versus rescaled $\tau$, respectively, with $\chi_{\nu}^2=1.668$ and $0.963$ demonstrating the collapse quality in the nonequilibrium scaling region to the right of the dashed lines.}
    \label{fig:tV}
\end{figure}

Here, we present the procedure for unraveling the critical properties via the short-imaginary-time scaling of the observable as dictated in \Eq{scaling}. In conventional PQMC, long imaginary-time evolution is typically required~\cite{SM,Sorella1989EPL,Sorella1991IJMPB,AssaadReview}, resulting in the severe sign problem exponentially increasing with $\tau$, as shown in the inset of Fig.~\ref{fig:SLAC}\subref{a}. 
We overcome this critical difficulty by leveraging short-imaginary-time critical relaxation dynamics, with the DSM state chosen as the initial state for the following demonstration.
To determine the critical point via \Eq{scaling}, we consider observable $O$ as the dimensionless correlation length ratio $R_{\rm FM}$ for the FM order (See SM Sec.~\ref{sec:S2} for the detailed definition~\cite{SM}). 
We take a short imaginary-time length $\tau=0.3L^z$, where \Eq{scaling} reduces to $R_{\rm FM}=f_R(gL^{1/\nu})$. Here the dynamical exponent is $z=1$ for the Dirac QCP, and our framework can also determine $z$ when unknown (see SM Sec.~\ref{sec:S2}~\cite{SM}), without any prior knowledge.
We show the numerical results in Fig.~\ref{fig:SLAC}\subref{b}, in which the crossing point of $R_{\rm FM}$ versus $U$ for different $L$ precisely reveals a quantum phase transition from DSM to FM phase occurring at $U_c = 7.220(37)$. 
This analysis fully aligns with the standard procedure of finite-size scaling, given the fixed relation $\tau = 0.3 L^z$. 
We then perform data collapse analysis of $R_{\rm FM}$ for different $L$ versus $(U-U_c) L^{1/\nu}$ and achieve accurate critical exponent $1/\nu = 1.18(3)$, as shown in Fig.~\ref{fig:SLAC}\subref{c}. 
Furthermore, fixing $U=U_c$, we calculate the imaginary-time evolution of FM structure order $S_{\rm FM}$ and fermion correlation $G_{\rm f}$ (defined in SM Sec.~\ref{sec:S2}~\cite{SM}), whose scaling dimensions are $(1+\eta_\phi)$ and $\eta_\psi$, respectively~\cite{Vaezi2022prl,yu2023dirac}. By making the rescaled curves of $S_{\rm FM}L^{(1+\eta_\phi)}$ and $G_{\rm f}L^{\eta_\psi}$ versus $\tau L^{-z}$ for different $L$ collapse according to \Eq{scaling} as shown in Figs.~\ref{fig:SLAC}\subref{d} and \ref{fig:SLAC}\subref{e}, we achieve the anomalous dimensions of the order parameter boson $\eta_{\phi} = 0.36(3)$ and the Dirac fermion $\eta_{\psi} = 0.134(3)$, respectively. 

The choice of the initial state and the imaginary-time length are two pivotal handles in our framework. Within the nonequilibrium scaling region near the QCP, the universal scaling form \Eq{scaling}, as well as the resulting critical point and exponents, are independent of these choices, while the severity of the sign problem is sensitive to them. Flexible choices of the initial state and imaginary-time length not only help alleviate the sign problem to the greatest extent, but also provide a self-consistency check for the accuracy of the extracted critical point and exponents. 
In the demonstration shown in Figs.~\ref{fig:SLAC}\subref{b} and \ref{fig:SLAC}\subref{c}, we used the DSM initial state with $\tau=0.3L^z$ to locate the critical point. The same result can be reproduced by changing $\tau$ to $\tau=0.5L^z$ or by switching the initial state to the FM state~\cite{SM}. Moreover, the scaling collapses in Figs.~\ref{fig:SLAC}\subref{d} and \ref{fig:SLAC}\subref{e} confirm that the system has entered the nonequilibrium scaling region. The details of self-consistency checks for different initial states and $\tau$ are provided in SM Sec.~\ref{Sec:S2-B}~\cite{SM}. Beyond the self-consistency checks, our estimates of the critical point and exponents are also approximately consistent with previous results from Gutzwiller QMC~\cite{Vaezi2022prl} and functional renormalization group~\cite{Vacca2015PRD}. 
The substantial mitigation of the sign problem enables us to simulate larger system sizes up to $L=23$, yielding reliable results (a systematic discussion is given in SM Sec.~\ref{sec:S1-E}~\cite{SM}). Consequently, we unambiguously demonstrate that short-imaginary-time dynamics via QMC simulation offers a powerful approach to determine the phase transition point and critical exponents in sign-problematic strongly interacting models in a self-consistent, accurate, and unbiased manner, without relying on any prior knowledge.

\pdfbookmark[2]{Spinless t-V model}{sec4}
\noindent {\it Spinless $t$-$V$ model} --- \\
To further demonstrate the framework of our approach, we study another typical interacting Dirac-fermion model, termed as honeycomb spinless $t$-$V$ model, with the Hamiltonian: ~\cite{huffman2014prb,Wang2014njp,li2015njp,Wessel2016prb}:
\begin{equation}
 H=-t\sum_{\langle ij\rangle}c_{i}^\dagger c_{j}+V\sum_{\langle ij\rangle} \left({n_{i}-\frac{1}{2}}\right) \left({n_{j}-\frac{1}{2}}\right) \label{Hamiltonian1},
\end{equation}
where $t$ is nearest-neighbor (NN) hopping amplitude and $V$ denotes the strength of NN density interaction. The appearance of the sign problem in \Eq{Hamiltonian1} depends on the channel of Hubbard-Strotonovich (H-S) transformation. Hence, the model provides a genuine platform to confirm the accuracy and feasibility of our approach to unravel QCP in the presence of the sign problem.  Previous sign-free QMC studies, with the H-S transformation in the hopping channel, reveal the critical properties of the QCP separating DSM and charge-density-wave (CDW) phases~\cite{li2015njp}. Here, we decouple the interaction in the sign-problematic density channel~\cite{li2022asymptotic} and study the QCP via short-imaginary-time relaxation dynamics.

\begin{figure*}[t]
    \centering
    \includegraphics[width=\linewidth]{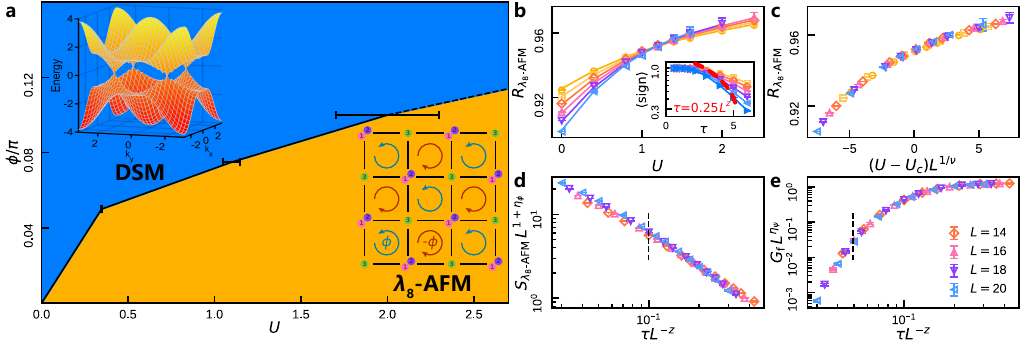}
    \caption{\textbf{Phase diagram and quantum criticality in $\rm SU(3)$ Hubbard model detected by short-imaginary-time dynamics with $\lambda_8$-AFM initial state.} \textbf{a}, Phase diagram determined via short-imaginary-time dynamics. Insets show the energy spectra of DSM state (upper left) and sketch of $\lambda_8$-AFM order in which a fermion with one flavor (green ``3") is situated at one sublattice and double fermions with the other two flavors (pink ``1" and violet ``2") are situated at the other sublattice (lower right). 
    \textbf{b}-\textbf{c}, Critical point $U_c=1.10(5)$ for $\phi=0.075\pi$ and $1/\nu=0.68(5)$ determined via curves of $R_{\lambda_8\text{-AFM}}$ versus $U$ for different $L$ at $\tau=0.25L^z$, where $z=1$ and $\chi_{\nu}^2=1.239$ for the collapse quality. Shown in Inset of {\bf b} is the evolution of average sign with red dashed curve marks $\tau=0.25L^z$. 
    \textbf{d}-\textbf{e}, $\eta_\phi = 0.55(5)$ and $\eta_\psi = 0.15(3)$ determined via the scaling collapse of evolution curves of $S_{\lambda_8\text{-AFM}}$ and $G_{\rm f}$, respectively, with $\chi_{\nu}^2=1.382$ and $\chi_{\nu}^2=0.560$ demonstrating the collapse quality in the nonequilibrium scaling region to the right of the dashed lines.}
    \label{fig:SU3}
\end{figure*}

The procedure is the same as in the previous section, but here we demonstrate the case with the CDW initial state. The consistency check with the DSM initial state is provided in SM Sec.~\ref{sec:S3}~\cite{SM}.
We fix $\tau=0.3 L^{z}$, and determine QCP from the crossing point of curves of correlation-length ratio $R_{\rm CDW}$ versus $V$ ($R_{\rm CDW}$ is defined in SM Sec.~\ref{sec:S3}~\cite{SM}), as displayed in Fig.~\ref{fig:tV}\subref{a}, giving rise to $V_c = 1.35(1)$. The scaling collapse analysis determines the value of $\nu$ as $1/\nu = 1.30(18)$. Additionally, the anomalous dimensions $\eta_\phi=0.49(5)$ and $\eta_\psi=0.073(4)$ are obtained from the evolution of the CDW structure factor $S_{\rm CDW}$ and the fermion correlation $G_{\rm f}$ (defined in SM Sec.~\ref{sec:S3}~\cite{SM}), respectively, at $V_c$. Both $V_c$ and $\eta_\phi$ are consistent with previous numerical results~\cite{Wang2014njp,li2015njp,Vacca2015PRD}. Notably, $\eta_\psi$ is numerically determined for the first time, aligning consistently with the previous results of the functional renormalization group~\cite{Vacca2015PRD}. 
These results for model~(\ref{Hamiltonian1}) further establish the feasibility of the approach based on short-imaginary-time dynamics in accessing the critical properties, even in the presence of the sign problem. The ability to select different initial states also offers valuable benchmarks, thereby further solidifying the reliability of our approach.

\pdfbookmark[2]{SU(3) Hubbard model}{sec5}
\noindent {\it SU(3) Hubbard model} --- \\
Building on the successful application of our method to previous models, we now apply it to systematically investigate the unexplored sign-problematic systems. Our particular interest lies in identifying an exotic Dirac QCP belonging to a novel universality class. To this end, we consider $\rm SU(3)$ Hubbard model on square lattice with staggered magnetic flux, described by the following Hamiltonian:
\bea
H = -\sum_{\avg{ij}\alpha} t_{ij} c^\dagger_{i\alpha} c_{j\alpha} + \frac{U}{2} \sum_i \left(\sum_{\alpha} n_{i\alpha}-\frac{3}{2}\right)^2,
\label{Ham3}
\eea
where $\alpha = 1$, $2$, $3$ is the flavor index of fermion, $U$ is repulsive Hubbard interaction strength, and $t_{ij}= t e^{i \theta_{ij}}$, in which $t=1$ is set as energy unit. As illustrated in Fig.~\ref{fig:SU3}\subref{a}, the magnetic flux in each plaquette is $\sum_{\square} \theta_{ij} = (-1)^{i_x+i_y}\phi$. For $\phi=0$ cases, recent constrained-path QMC and finite-temperature determinant QMC studies have investigated the model at 1/3 filling~\cite{Feng2023prr,Ibarra-Garcia-Padilla2023pra}. For non-zero $\phi$, the energy dispersion of \Eq{Ham3} in the non-interacting limit features two Dirac points located at momenta $(\pm \frac{\pi}{2},\pm \frac{\pi}{2})$, as shown in the inset of Fig.~\ref{fig:SU3}\subref{a}. We fix our simulation at half filling such that the Fermi level is located at the Dirac points. The model of \Eq{Ham3} respects $\rm SU(3)\times Z_2$ symmetry, where $\rm SU(3)$ is the rotation symmetry in flavor space and $\rm Z_2$ is the sublattice symmetry which exchanges the two sublattices on the staggered-flux square lattice with each unit cell containing two sites.

The $\rm SU(N)$ symmetry plays overarching roles in modern physics. For instance, the $\rm SU(3)$ symmetry lays the foundation for strong interaction between quarks~\cite{weinbergBook}. Recently, interacting fermionic models with $\rm SU(N)$ symmetry have been realized in optical lattice~\cite{Takahashi2018prl,Takahashi2022np}. Extensive theoretical and numerical efforts, including QMC simulations, are devoted to understanding the exotic phenomena arising from the interplay between strong interaction, Dirac fermions and multi-flavor physics~\cite{Assaad2013prl,wu2016prb,He2016PRB,li2017nc,Wang2023PRR}. However, despite its fundamental importance, unbiased numerical studies on $\rm SU(N)$-symmetric model with repulsive interaction for odd $N$ are scarce, primarily due to the notorious sign problem. Hence, we implement our approach to preempt the sign problem and unravel the quantum phases and exotic QCP in \Eq{Ham3}, a paradigmatic interacting model for odd $N$ $\rm SU(N)$ symmetry. 

Before embarking on QMC simulation, we perform a mean-field calculation to detect the feature of the ground state schematically. The mean-field calculation shows that an antiferromagnetic (AFM) order is favored by the strong Hubbard interaction~\cite{SM}. The order parameter is characterized in terms of the generators of $\rm SU(3)$ symmetry group, which is expressed by the eight Gell-Mann matrices in flavor space for convenience~\cite{Gell-Mann1962prl} (The details of the $\rm SU(3)$ algebra are introduced in SM~\cite{SM}). In stark contrast to the case of $\rm SU(2)$, only one Gell-Mann matrix is full-rank without zero eigenvalue, dubbed as $\lambda_8 = \rm{diag}(\frac{1}{\sqrt{3}},\frac{1}{\sqrt{3}},-\frac{2}{\sqrt{3}})$. Consequently, for the AFM order in $\rm SU(3)$ fermions, only the order parameter associated with $\lambda_8$ is the mass term fully opening spectral gap in Dirac fermions, which is expressed as $m_{\lambda_8\text{-AFM}} = \frac{1}{L^2}\sum_{i,\alpha,\beta} c^\dagger_{i\alpha} \lambda_8^{\alpha\beta} c_{i\beta} (-1)^{i_x+i_y}$. The mean-field calculation also confirms that such AFM order is energetically favorable in model \Eq{Ham3}~\cite{SM}.

In the $\lambda_8$-AFM ordered phase, the $\rm Z_2$ sublattice symmetry is broken and the $\rm SU(3)$ symmetry is broken into $\rm SU(2)\times U(1)$ (See SM Sec.~\ref{sec:S4-D} for details~\cite{SM}), as illustrated in Fig.~\ref{fig:SU3}\subref{a}.
This order parameter transforms under the adjoint representation of SU(3), spanning the vacuum manifold $\frac{\text{SU(3)}\times \text{Z}_2}{\text{SU(2)}\times \text{U(1)}}$. Such symmetry breaking defines a new unconventional Gross-Neveu QCP, where the order parameter symmetry is SU(3), in contrast to the O($N$) order parameters of the conventional Gross-Neveu universality classes such as chiral Ising, chiral XY, and chiral Heisenberg.

To reveal the critical properties, we perform short-time QMC simulation starting from a fully ordered state at $\phi=0.075\pi$ and $\tau L^{-z}=0.25$ ($z=1$). The results of correlation length ratio $R_{\lambda_8\text{-AFM}}$ for $\lambda_8$-AFM order (defined in SM Sec.~\ref{sec:S4}~\cite{SM}) are shown in Fig.~\ref{fig:SU3}\subref{b} and the crossing points dictate the phase transition point from DSM to $\lambda_8$-AFM phase occurring at $U_c = 1.10(5)$. We also compute the correlation-length ratios for other types of AFM order instead of $\lambda_8$, confirming that only $\lambda_8$-AFM long-range order is present in the ground state~\cite{SM}. With varying magnetic flux $\phi$, we apply the same procedure to map out the ground-state phase diagram of the \Eq{Ham3}, as shown in Fig.~\ref{fig:SU3}\subref{a}, in which $U_c$ increases with $\phi$. 

After accessing the phase diagram, we investigate the critical properties. The data collapse analysis of $R_{\lambda_8\text{-AFM}}$ versus $(U-U_c) L^{1/\nu}$ in the regime close to QCP with fixed $\phi=0.075$ gives $1/\nu = 0.68(5)$, as shown in Fig.~\ref{fig:SU3}\subref{c}. Moreover, at $U_c$, the short-imaginary-time dynamics of the structure factor $S_{\lambda_8\text{-AFM}}$ and the fermion correlation function $G_{\rm f}$  (defined in SM Sec.~\ref{sec:S4}~\cite{SM}) yield $\eta_\phi = 0.55(5)$ and $\eta_\psi = 0.15(3)$, as presented in Fig.~\ref{fig:SU3}\subref{d} and \ref{fig:SU3}\subref{e}, respectively. To confirm the universality of this phase transition, we also evaluate the critical exponents for other values of $\phi$, as presented in SM~\cite{SM}, and they are mutually consistent within error bars.
Remarkably, the critical exponents found here are numerically distinct from those of any conventional Gross-Neveu universality class~\cite{Rosenstein1993PLB,Herbut2014prb} (for comparison, see SM~\cite{SM}), providing conclusive evidence that this transition belongs to a new universality class.

\pdfbookmark[1]{Discussion}{Discussion}
\noindent\textcolor{YinshuaiBlue}{\textbf{DISCUSSION}}\\
In summary, we introduce an innovative theoretical framework unraveling ground-state properties, with a particular focus on quantum criticality, in quantum many-body systems. Our method effectively circumvents the notorious sign problem in QMC by leveraging the short-stage imaginary-time relaxation dynamics. We provide compelling evidence of its accuracy and efficiency across several strongly interacting models, showcasing a dramatic reduction in computational cost compared to conventional equilibrium QMC. For example, the estimate time for the $\rm SU(3)$ Hubbard model with $L=10$ is billions of times shorter with our approach than with equilibrium PQMC~\cite{SM}. Furthermore, the inherent flexibility in choosing initial states, coupled with a robust scaling relation, provides a crucial benchmark for ensuring accuracy within our framework. Besides, this approach can distinguish between first-order and continuous phase transitions~\cite{SM}. This work, therefore, establishes a powerful strategy for the unbiased numerical study of the exotic quantum phases and quantum phase transitions in a broad class of previously computationally inaccessible systems.

More intriguingly, we leverage our newly developed approach to make a pioneering investigation into the $\rm SU(3)$ repulsive Hubbard model hosting Dirac fermions.  It represents the \textit{first time} that the ground-state phase diagram and the QCP in the $\rm SU(N)$ repulsive Hubbard model with odd $N$ have been unveiled through unbiased numerical approaches. We identify the first example of Gross-Neveu QCP that fundamentally differs from the conventional ones typically confined to the chiral O($N$) universality classes~\cite{SM}. 
Moreover, the critical properties are potentially detectable in the synthetic quantum simulators such as optical lattice of cold atoms~\cite{Takahashi2018prl,Takahashi2022np}. Hence, our findings on the $\rm SU(3)$ Hubbard model not only establish a significant novel direction in modern statistics physics for exploring the fermionic QCP but also offer critical theoretical guidance for the forthcoming experimental exploration  of exotic physics in strongly correlated $\rm SU(N)$ fermionic models. 

Our innovative approach presents a powerfully generalizable framework, extending readily to other fermionic QCP, including the notoriously challenging metallic QCP involving Fermi surfaces, whose critical properties are of immense theoretical and experimental interest but have remained largely inscrutable. Moreover, the short-time nonequilibrium QMC is also applicable to sign-problematic bosonic models, opening entirely new avenues.  We carefully delineate the boundaries of our method: it is not a fully generalized solution to the sign problem.  Arbitrarily short evolution times retain excessive memory of the initial state, preventing evolution into the desired nonequilibrium scaling regime (see SM Sec. \ref{sec:S1}~\cite{SM} for detailed discussions). This means a mild sign problem ensuring accurate results within the short-time nonequilibrium scaling regime cannot be universally guaranteed for every model. For instance, in the SU(3) Hubbard model with large staggered magnetic flux, the sign problem remains too severe for highly accurate results near the QCP, even in the short-time stage. Despite these limitations, the framework's flexibility allows for future efficiency enhancements, such as through improvement of initial state and optimization of H-S channel~\cite{Yao2022prb,Chang2024PRB}. This work is not merely an incremental improvement; it delivers a novel pathway toward establishing a general and unbiased numerical strategy for unraveling the mysteries of the ground-state phase diagram and QCP in strongly correlated systems, a long-sought goal in the field.

\bibliographystyle{apsrev4-2}
\bibliography{ref}

\noindent\textbf{Acknowledgments:} We acknowledge helpful discussions with Yiwen Pan, Chengfeng Cai, Shao-Hang Shi, and Chang-Yu Shen.

\noindent\textbf{Funding Statement:}  
Yin-Kai Yu, Zhi-Xuan Li, and Shuai Yin are supported by the National Natural Science Foundation of China (Grants No. 12222515 and 12075324).  
Zi-Xiang Li is supported by the National Natural Science Foundation of China under Grant No. 12347107.  
Shuai Yin is also supported by the Science and Technology Projects in Guangdong Province (Grant No. 2021QN02X561) and in Guangzhou City (Grant No. 2025A04J5408).

\noindent\textbf{Author Contributions Statement:} Shuai Yin and Zi-Xiang Li conceived and supervised the project. Yin-Kai Yu and Zhi-Xuan Li carried out the numerical calculations and data analysis. All authors participated in discussions and contributed to writing the manuscript. 

\noindent\textbf{Competing Interests Statement:}  
The authors declare they have no competing interest.

\noindent\textbf{Data \& Materials Availability Statement:}  
All raw data underlying the results presented in the main text and Supplementary Materials are archived in the Dryad Digital Repository (\href{https://doi.org/10.5061/dryad.dncjsxmch}{DOI: 10.5061/dryad.dncjsxmch}) under the CC0 Public Domain Dedication.

\onecolumngrid
\newpage
\widetext
\thispagestyle{empty}

\setcounter{equation}{0}
\setcounter{figure}{0}
\setcounter{table}{0}
\renewcommand{\theequation}{S\arabic{equation}}
\renewcommand{\thefigure}{S\arabic{figure}}
\renewcommand{\thetable}{S\arabic{table}}
\renewcommand{\thesection}{S\arabic{section}}

\renewcommand\floatpagefraction{0.9}
\renewcommand\textfraction{0.1}

\pdfbookmark[0]{Supplementary Materials}{SM}
\begin{center}
    \vspace{3em}
    {\Large\textbf{Supplementary Materials for}}\\
    \vspace{1em}
    {\large\textbf{Preempting Fermion Sign Problem: Unveiling Quantum Criticality through Nonequilibrium Dynamics in Imaginary Time}}\\
    \vspace{0.5em}
\end{center}

\tableofcontents

\newpage
\section{Details of nonequilibrium imaginary-time dynamics in PQMC simulation}
\label{sec:S1}

    \subsection{Imaginary-time relaxation dynamics simulated by PQMC}
        
         We focus on the imaginary-time relaxation dynamics from a fully ordered or Dirac semimetal initial state $\ket{\psi_0}$. The initial state is prepared by solving $H_0 \ket{\psi_0} = E_0 \ket{\psi_0}$, where $H_0$ is the initial Hamiltonian and $E_0$ is the ground state energy. With these initial states, the evolution of the observable $O(\tau)$ is given by
        \begin{equation}
            \braket{O(\tau)} = 
            \frac{
                    \bra{\psi_0}\mathrm{e}^{-\frac{\tau}{2}H}~O~\mathrm{e}^{-\frac{\tau}{2}H}\ket{\psi_0}
                }{
                    \bra{\psi_0}\mathrm{e}^{-\tau H}\ket{\psi_0}
                }. \label{s1}
        \end{equation}
        As $\tau\to\infty$, $\mathrm{e}^{-\frac{\tau}{2}H}$ projects the system onto the ground state of $H$.
        
        The imaginary-time relaxation dynamics can be simulated via the projector quantum Monte Carlo (PQMC)~\cite{Sorella1989EPL,Sorella1991IJMPB,AssaadReview}. In conventional PQMC studies, a sufficiently large $\tau$ (usually $\tau$ should be several times as large as $L$) is needed to ensure that the ground state is obtained. Then the physical qualities are calculated in the ground state. 
        
        In contrast, in our work, we focus on the short-time stage of the imaginary-time relaxation process, where $\tau$ does not need to be large compared to $L$. In this regime, the system remains in a nonequilibrium state rather than the ground state This stage is usually discarded in conventional PQMC studies. In our previous work, we systematically and carefully investigated the imaginary-time nonequilibrium dynamics of Dirac fermion systems~\cite{yu2023dirac}, and found that although this setup goes beyond the conventional PQMC framework, choosing a small $\tau$ only introduces some dependence on the initial state, while the ground-state information, particularly universal quantum criticality, still manifests in the nonequilibrium dynamics. With a properly chosen initial state, one can fix $\tau / L^z$ to a small value and still observe quantum phase transitions via standard finite-size scaling analysis, even in this nonequilibrium regime.
        
        A similar practice has also appeared earlier in the context of finite-temperature QMC (FTQMC). In FTQMC, the expectation value of an observable is given by $\braket{O} = \frac{1}{Z} \text{Tr}[\mathrm{e}^{-\beta(H - \mu N)} O]$, with $Z = \text{Tr}[\mathrm{e}^{-\beta(H - \mu N)}]$~\cite{Hirsch1981PRL,Hirsch1985PRB,White1989prb}. Here, $\beta = 1/k_\text{B} T$ is the inverse temperature, and it corresponds to the imaginary-time extent in the Monte Carlo simulation. Unlike zero-temperature PQMC, FTQMC involves both quantum and thermal fluctuations. When FTQMC is used to study zero-temperature quantum phase transitions, it is common to fix $\beta / L^z$ to a finite value (typically $\beta = L^z$) to obtain temperature-independent finite-size scaling. There are also studies using smaller values, such as $\beta = \frac{1}{4} L^z$~\cite{Sorensen1992prl}.
        
        Although both $\beta$ and $\tau$ represent the imaginary-time length in QMC simulations, they carry distinct physical meanings. A shorter $\beta$ in FTQMC corresponds to a higher-temperature thermal equilibrium state, and it is widely accepted that $\beta / L^z$ can be used as a dimensionless scaling factor. In contrast, a shorter $\tau$ in zero-temperature PQMC corresponds to a nonequilibrium state and has rarely been considered outside the context of imaginary-time nonequilibrium dynamics~\cite{Yins2014prb,Shu2017prb,yu2023dirac}. Here, based on our understanding of imaginary-time nonequilibrium dynamics, we leverage this idea to address the fermion sign problem in zero-temperature PQMC simulation. Moreover, compared to conventional PQMC and FTQMC approaches, our framework further alleviates the sign problem by flexibly choosing appropriate initial states, while the consistency of results obtained from different initial states provides a useful means of self-consistency check.

    \subsection{Hubbard-Stratonovich transformation}

        In the PQMC simulations, the interaction terms in the form of four-fermion operators should be decoupled via Hubbard-Stratonovich (HS) transformation. The implementation of PQMC begins with performing the Trotter decomposition to discretize the imaginary-time evolution operator into ${M=\tau/\Delta\tau}$ (where ${M}$ is an integer) time slices, i.e.,
        \begin{equation}
            \mathrm{e}^{-\tau H} = \prod^M_{m=1} \left[\mathrm{e}^{-\Delta\tau H_t} \mathrm{e}^{-\Delta\tau H_U} + \mathcal{O}\left(\Delta\tau^2\right)\right],
        \end{equation}
        where $H_t$ and $H_U$ represent the free fermion hopping term and the interaction term in the Hamiltonian, respectively. Then, we use the HS transformation on $H_U$ to decouple the fermion-fermion interaction into interactions between non-interacting fermions and auxiliary fields.
        
        For the single-Dirac-fermion Hubbard model, we use the following HS transformation:
        \begin{equation}
            \expup{-\Delta \tau U \left(n_{i\uparrow}-\frac{1}{2}\right)\left(n_{i\downarrow}-\frac{1}{2}\right)}
            = \frac{1}{2} \expup{-\frac{\Delta \tau U}{4}} \sum_{s_i=\pm 1} \expup{\lambda s_i (c_{i\uparrow}^\dagger c_{i\downarrow}+c_{i\downarrow}^\dagger c_{i\uparrow})}.
        \end{equation}
        where $\cosh \lambda=\expup{\frac{\Delta \tau U}{2}}$. For $U>0$ (Hubbard repulsive interaction), the sign problem arises for all channels. We here choose the $\sigma_x$ channel to mitigate the sign problem~\cite{Vaezi2022prl}.
        This choice also affects the structure of the fermion determinant. Specifically, due to the spin-flipping nature of the free-fermion hopping term in the single-Dirac-fermion Hubbard model, the $\sigma_x$-channel decoupling leads to an auxiliary-field Hamiltonian that only has off-diagonal blocks. As a result, the total fermion determinant can be factorized into the product of two smaller determinants: $\det(M) \propto \det(M_{\uparrow\downarrow}) \det(M_{\downarrow\uparrow})$. 
       In contrast, decoupling in the $\sigma_z$ channel does not permit this factorization. This results in a single determinant of twice the dimension, which typically exacerbates the sign problem.
        
        For the spinless $t$-$V$ model, previous studies have shown that at half-filling, the model can be decoupled into the hopping channel without the sign problem, which has been demonstrated from various perspectives \cite{li_solving_2015,li_majorana-time-reversal_2016,xiang2016prl,wang2015prl}. In our study, however, we use the HS transformation in a sign-problematic channel, namely the density channel:
        \begin{equation}
            \expup{-\Delta \tau V \left({n_{i}-\frac{1}{2}}\right) \left({n_{j}-\frac{1}{2}}\right) }
            = \frac{1}{2} \expup{-\frac{\Delta \tau V}{4}} \sum_{s_{ij}=\pm 1} \expup{\lambda s_{ij} \left(n_i-n_j\right)}.
        \end{equation}
        where $\cosh \lambda=\expup{\frac{\Delta \tau V}{2}}$. Despite the presence of sign problem, the numerical results we obtained are consistent with the previously established results, demonstrating the reliability of our new method.

        For the $\rm SU(3)$ repulsive Hubbard model, we use the following HS transformation:
        \begin{equation}
            \expup{\Delta \tau \frac{U}{2} \left(n_{i\alpha}-n_{i\beta}\right)^2}
            = \frac{1}{2} \sum_{s_i=\pm 1} \expup{\lambda s_i \left(n_{i\alpha}-n_{i\beta}\right)}.
        \end{equation}
        where $\cosh \lambda=\expup{\frac{\Delta \tau U}{2}}$. For all the known algorithms, the $\rm SU(3)$ repulsive Hubbard model is sign-problematic for any decoupling channel in HS transformation.  

    \subsection{Sign problem in PQMC}
        
        Through the HS transformation, the Hamiltonian can be converted into a quadratic effective form of fermionic operators that depends on the spacetime configuration of the auxiliary fields. The partition function can then be expressed as a sum of configuration weight $w(\mathrm{c})$, i.e., $Z=\sum_{\mathrm{c}} w (\mathrm{c})$. These weights are given by the determinant of the effective Hamiltonian of fermions~\cite{AssaadReview}.
        
        In PQMC simulations, we sample the space-time dependent configuration of the auxiliary field. For sign-free models, the sampling probability is proportional to the configuration weight $w(\mathrm{c})$. However, for the sign-problematic models, the configuration weight $w(\mathrm{c})$ is not positive definite, so it cannot be used directly as the sampling probability. Instead, the absolute value \nota{|w (\mathrm{c})|} is used as the sampling probability, and the observables are computed as follows \cite{li2019review, Troyer2005sign}:
        \begin{equation}
            \braket{O}=\frac{\sum_{\mathrm{c}} w(\mathrm{c}) O(\mathrm{c})}{\sum_{\mathrm{c}} w(\mathrm{c}) }
            =\frac{\sum_{\mathrm{c}} |w(\mathrm{c})| \mathrm{sign}(\mathrm{c}) O(\mathrm{c})/\sum_{\mathrm{c}} |w(\mathrm{c})| }{\sum_{\mathrm{c}} |w(\mathrm{c})| \mathrm{sign}(\mathrm{c})/\sum_{\mathrm{c}} |w(\mathrm{c})|}
            =\frac{\braket{O\,\mathrm{sign}}_{|w|}}{\braket{\mathrm{sign}}_{|w|}}.
            \label{eq:observable}
        \end{equation}
        Here, we have used
        \begin{equation}
            \braket{~\Box ~}_{\abs{w}}=\frac{\sum_{\mathrm{c}} ~\Box ~ |w(\mathrm{c})| }{\sum_{\mathrm{c}} |w(\mathrm{c})|},
        \end{equation}
        to denote the expectation value obtained using $|w (\mathrm{c})|$ as the sampling probability. The sign problem introduces a cost that the average sign $\braket{\mathrm{sign}}_{|w|}$ tends to zero due to the frequent cancellation of positive and negative weights across different configurations, leading to the consequence that \eqref{eq:observable} becomes a ratio between two tiny numbers. This is numerically unstable and introduces significant statistical errors. Specifically, it is proven that the error generally follows \cite{Troyer2005sign}:
        \begin{equation}
            \Delta \braket{O}\propto\frac{1}{\braket{\mathrm{sign}}_{|w|}}\propto\expup{\tau N\Delta f},
            \label{eq:sign_err}
        \end{equation}
        where $\Delta f$ denotes the difference in the free energy density between the actual fermionic system and its corresponding bosonic system. The exponential dependence of the error amplification factor on the imaginary time $\tau$ and the number of particles $N$ means that QMC requires exponentially long computational times to achieve controllable statistical errors when solving ground-state problems of quantum systems in the thermodynamic limit.
        
        In subsequent analyses, we use the average sign $\braket{\mathrm{sign}}_{|w|}$ to measure the severity of the sign problem. The lower value of the average sign $\braket{\mathrm{sign}}_{|w|}$ indicates a more severe sign problem.

    \subsection{Applicability and limitations of the nonequilibrium approach}

        The proposed nonequilibrium method offers an efficient pathway for exploring ground-state properties by circumventing the sign problem in our model and significantly lowering computational cost. This makes it highly effective for mapping the phase diagram and extracting critical properties of the quantum phase transition. However, it is important to emphasize that this is not a universal solution for all models plagued by the sign problem. Here, we discuss the applicability and limitations of our approach. The nonequilibrium approach is subject to two key limitations that constrain its applicability:
        \begin{enumerate}
            \item One cannot take the imaginary time arbitrarily short in order to ensure that the sign problem is mild for all models. If the evolution time is too short, the system retains too much memory of the initial state, and the scaling theory breaks down. For instance, this breakdown of scaling theory occurs when the characteristic length scale of the evolution, $\xi_l \propto \tau^{1/z}$, becomes comparable to the system's ultraviolet (UV) cutoff---in our case, the lattice spacing (set to unity). Under this condition, the behavior of observables is dominated by non-universal, high-energy physics, rendering the scaling analysis invalid.
            \item This method is applicable to the continuous quantum phase transitions. This method is applicable for studying continuous quantum phase transitions. Our approach can be applied to determine the ground-state phase diagram and to identify and locate possible QCPs. If a QCP exists, our method can accurately determine the critical exponents based on the scaling theory of continuous transitions. On the other hand, if the calculation reveals that the transition does not exhibit scaling behavior, it implies a first-order transition, for which there is currently no well-established scaling analysis framework to use. In addition, our method is not applicable deep in the ordered phase, far away from the transition point.
        \end{enumerate}

        We begin by elaborating on the first limitation, and provide a framework for understanding when the method is effective and when it fails.
        In general, fixing a system size $L$ for discussion, the effectiveness of our method depends on three characteristic imaginary-time scales:  
        (1) the minimum time $\tau_{\text{neq.}}$ at which the nonequilibrium scaling theory becomes valid;  
        (2) the minimum time $\tau_{\text{eq.}}$ required for conventional PQMC to project onto the ground state; and  
        (3) the maximum time $\tau_{\text{sign}}$ before the sign problem becomes too severe to obtain reliable results. 
        Empirically, $\tau_{\text{eq.}}$ varies slightly between models but is usually around two to three times the linear system size $L$. The time $\tau_{\text{neq.}}$ required for short-time scaling to apply is typically one or two orders of magnitude smaller than $\tau_{\text{eq.}}$. In contrast, $\tau_{\text{sign}}$ can differ drastically from model to model and is the dominant factor that determines whether our approach will be effective.
        This leads to three possible scenarios:
        \begin{enumerate}[label=\arabic*.]
            \item $\tau_{\text{neq.}} < \tau_{\text{eq.}} < \tau_{\text{sign}}$: the sign problem is mild, and equilibrium methods already work well. Our nonequilibrium method only improves computational efficiency.
            \item $\tau_{\text{neq.}} < \tau_{\text{sign}} < \tau_{\text{eq.}}$: equilibrium methods fail due to the sign problem, but the nonequilibrium method still works. This is the type of situation our work is focused on.
            \item $\tau_{\text{sign}} < \tau_{\text{neq.}} < \tau_{\text{eq.}}$: the sign problem is too severe even at short times. In this case, the nonequilibrium method also fails. 
        \end{enumerate}
        Because the sign problem behaves differently across models, $\tau_{\text{sign}}$ varies accordingly. Based on this classification, we can view sign-problematic strongly correlated models as falling into three broad categories. The general argument that a QCP can be approached using sufficiently short imaginary-time simulations where the sign problem remains mild is valid only in models of type~1 and type~2. It does not hold in type~3 models, where the sign problem is already severe even at short times. Naturally, our method is not a fully general solution to the sign problem in all cases. What it does provide is a practical and effective approach for models of type~2, which were previously inaccessible due to severe sign problems in equilibrium simulations. With our method, such models become tractable. 

\begin{figure}[htbp]
            \centering
            \includegraphics{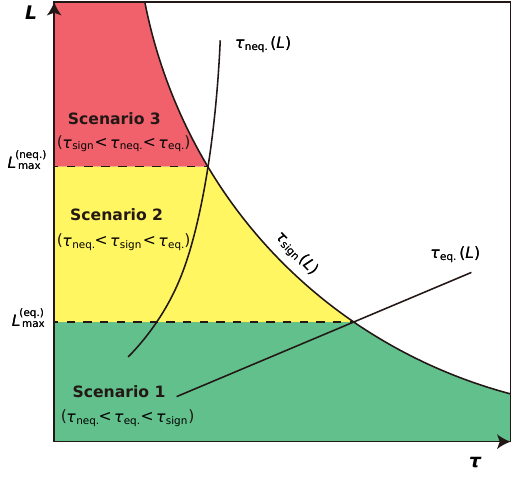}
            \caption{Schematic illustration of the three scenarios regarding the sign problem. The shaded region indicates where the sign problem is not severe and simulations are feasible. The green region corresponds to scenario 1 ($\tau_{\text{neq.}} < \tau_{\text{eq.}} < \tau_{\text{sign}}$), where equilibrium methods also work. The yellow region corresponds to scenario 2 ($\tau_{\text{neq.}} < \tau_{\text{sign}} < \tau_{\text{eq.}}$), where only the nonequilibrium method is feasible. The red region corresponds to scenario 3 ($\tau_{\text{sign}} < \tau_{\text{neq.}} < \tau_{\text{eq.}}$), where even the nonequilibrium approach fails. The boundaries define the maximum accessible system sizes $L_{\text{max}}^{(\text{eq.})}$ and $L_{\text{max}}^{(\text{neq.})}$ for each method.}
            \label{fig:draft}
        \end{figure}

        Within the same model and parameter set, $\tau_{\text{sign}}$ also depends on the system size $L$.  Typically, $\tau_{\text{sign}}$ decreases as $L$ increases. This is because the average sign scales exponentially with both the particle number---which grows with system size ($N \propto L^d$)---and $\tau$, as described by \eqref{eq:sign_err}~\cite{Troyer2005sign}, although exceptions have been noted~\cite{Scalettar2015prb,li2022asymptotic}.  Taking this size dependence into account, we can sketch a regime diagram in the $(L, \tau)$ plane with the three regimes described above, as illustrated in Fig.~\ref{fig:draft}. This diagram clearly demarcates the three previously described regimes and highlights a key advantage of the nonequilibrium method: it grants access to significantly larger system sizes than those achievable with conventional equilibrium QMC.
        
        The above classification only provides general criteria for when the nonequilibrium method is effective, while the quantitative determination of the nonequilibrium scaling regime in practical applications is illustrated in detail with procedures and examples in Sec.~\ref{Sec:S2-B}.

        Next, we discuss another boundary case of this nonequilibrium short-time method, i.e., the first-order phase transition. In the main text, we have already demonstrated that our nonequilibrium method can determine the existence of a continuous phase transition based on scaling forms.
        The key criteria are: 
        1) For a given $\tau L^{-z}$, the crossing points of the correlation-length ratio or Binder ratio curves converge to a single point as the system size increases (dimensionless quantities exhibit scale invariance at the critical point). 
        2) The nonequilibrium critical relaxation processes of physical quantities such as the structure factor and fermion correlation display scaling collapse.

        In fact, we are also able to identify a first-order phase transition during the short-time evolution through opposite characteristics: 1) The crossing points of the correlation-length ratio or Binder ratio curves do not converge to a single point. 2) The nonequilibrium critical relaxation processes of physical quantities such as the structure factor and fermion correlation cannot be well scaled to collapse. Additionally, 3) due to the coexistence of two phases at a first-order phase transition, the Binder ratio typically shows negative dips at the transition point, and the Monte Carlo sampling distribution of the structure factor exhibits a double peak. The emergence of negative dips in the results of Binder ratio is a hallmark of the first-order transition.

\begin{figure}[bt]
            \centering
            \includegraphics[width=\linewidth]{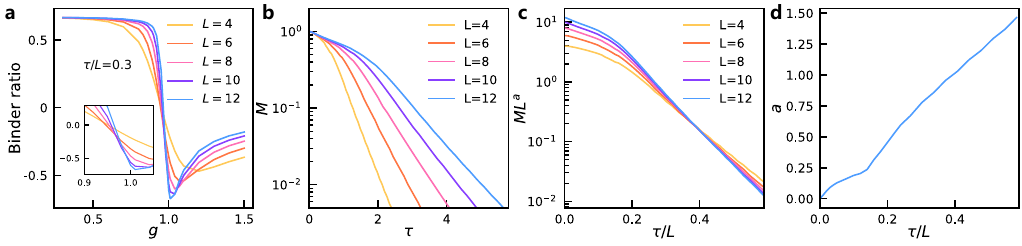}
            \caption{
                \textbf{First-order phase transition characteristics in the $q=6$ quantum Potts chain observed during short imaginary time evolution.}
                \textbf{a}, Binder ratio versus $g$ for different system sizes, where the curves do not intersect at a single point, and negative dips are observed. The emergence of negative dips in the Binder ratios is a hallmark of first-order transition. 
                \textbf{b}, Imaginary time relaxation of the order parameter $M$ starting from the ordered phase at $g=1$.
                \textbf{c}, Failure of scaling collapse for \textbf{b}, with $a=1.0$ as an example.
                \textbf{d}, Fitting $M \propto L^{-a}$ for fixed $\tau/L$, where the value of $a$ does not converge as $\tau/L$ increases.
            }
            \label{fig:FIG_Potts}
        \end{figure}

        As a typical example, we demonstrate how to identify the first-order phase transition in the $q=6$ quantum Potts chain during short imaginary time evolution. The Hamiltonian is given by:
        \begin{equation}
            H = -qJ \sum_{i} \sum_{m=0}^{q-1} P_i^{(m)} P_{i+1}^{(m)} 
            \;-\; h \sum_{i} \sum_{m\neq n} |m\rangle_i\langle n| \, ,
        \end{equation}
        where $P_i^{(m)} \equiv |m\rangle_i \langle m|$ is the projection operator on site $i$, and the tuning parameter is $g\equiv h/J$. For $g<1$, the ground state is in the ferromagnetic phase, and for $g>1$, it is in the paramagnetic phase. For the case of $q=6$, a first-order phase transition occurs at $g_c=1$. We use the time-evolving block decimation (TEBD) method to simulate the system's imaginary time evolution starting from an ordered ferromagnetic initial state (e.g., all sites choosed into $m=0$ state) and observe the order parameter $M$ and Binder ratio $R$:
        \begin{equation}
            M \equiv \frac{1}{L}\sum_{i=1}^L s_i \, ,\qquad R \equiv 1 - \frac{\langle M^4\rangle}{3\,\langle M^2\rangle^2} \, ,
        \end{equation}
        where $s_i \equiv \frac{q}{q-1} \left(P_i^{(0)} - \frac{1}{q}\right)$. Using the same procedure as in the main text, we take a short imaginary time $\tau/L=0.3$. The variation of Binder ratio $R$ with $g$ is shown in Fig.~\ref{fig:FIG_Potts}\subref{a}. The curves for different system sizes do not intersect at a single point, and there are distinct negative dips, which are characteristic of a first-order phase transition and can appear with short imaginary time evolution at $\tau/L=0.3$. Fig.~\ref{fig:FIG_Potts}\subref{b} shows the imaginary time relaxation of the order parameter $M$ starting from the ordered phase at $g=1$, which decays exponentially with $\tau$. Fig.~\ref{fig:FIG_Potts}\subref{c} demonstrates the failure of scaling collapse for $M$, using a scaling exponent $a=1.0$ as an example. Adjusting $a$ does not allow these curves to collapse or partially collapse. If we fix $\tau/L$ and perform scaling fitting according to the form $M \propto L^{-a}$, the results for $a$ are shown in Fig.~\ref{fig:FIG_Potts}\subref{d}, and they do not converge as $\tau/L$ increases. These results, which violate scaling forms, clearly exclude the possibility of a continuous phase transition here. This example illustrates that a first-order phase transition can be distinguished from a continuous phase transition even during short imaginary time evolution.
        However, for such cases, although the rough phase diagram can still be determined, the precise location of the transition point may not be accurately obtained within the short imaginary-time regime since the current scaling form no longer holds, and may instead require extrapolation from finite-time data.

    \subsection{Guiding principles for choosing the value of \nota{\tau/L^z} in simulations}
    \label{sec:S1-E}

        In the main text, we adopt the form $\tau = \text{(const.)} \times L^z$ to identify quantum critical points. We use coefficients $0.3$ in Figs.~\ref{fig:SLAC}\subref{b}–\subref{c} and \ref{fig:tV}\subref{a}–\subref{b}, and $0.25$ in Figs.~\ref{fig:SU3}\subref{b}–\subref{c}. When choosing these specific coefficients, we carefully balanced three considerations: the severity of the sign problem, the validity of the scaling theory, and the numerical precision of the estimated critical point. In general, the following guiding principles apply:
        \begin{enumerate}
            \item The severity of the sign problem is the most important factor. Usually, slightly increasing the value of $\tau/L^z$ makes the sign problem significantly worse, and vice versa. For example, if the coefficient is increased from $0.3$ to $0.6$, the sign problem becomes notably more severe, significantly reducing the accessible system size. From this guiding principle, $\tau/L^z$ should be chosen as small as possible.
            \item Another important factor is the validity of the scaling theory. If $\tau/L^z$ is too small, the nonequilibrium scaling theory may not hold well at currently accessible finite sizes. Specifically, for example, when it is taken below 0.1, from Figs.~\ref{fig:SLAC}\subref{d}, \ref{fig:tV}\subref{c}, and \ref{fig:SU3}\subref{d} one can see that the data collapse gradually starts to break down, meaning finite-size effects become significant. This condition is relatively flexible. As long as $\tau/L^z$ is above a certain threshold, the scaling holds well, and variations in the coefficient do not significantly affect the results. Values such as $0.25$, $0.3$, and $0.5$ are typically used in the literature~\cite{Yin2022prl,yu2023dirac}, where estimates of critical points at different $\tau/L^z$ values were found to agree within statistical error, providing a useful self-consistency check.
            \item Another somewhat subtle factor is the resolution of the critical point. Even within the valid scaling regime, choosing a smaller $\tau/L^z$ broadens the nonequilibrium critical region. This broadening manifests as a diminished dependence of the dimensionless correlation length ratio $R$ on system size $L$ over a wide range of the tuning parameter $U$ near $U_c$, making the crossing point of $R$ curves (such as in Fig.~\ref{fig:SLAC}\subref{b}) less sharp. This leads to reduced numerical precision in locating the critical point. Therefore, from this perspective, it is preferable to choose a somewhat larger $\tau/L^z$ to enhance resolution.
        \end{enumerate}
        In summary, when applying the nonequilibrium approach to determine quantum critical points, one must balance these considerations within the valid scaling regime. The goal is to choose a value of $\tau/L^z$ that optimizes both precision and accuracy. In our experience, the severity of the sign problem is often the most decisive factor.

\section{More details for the single-Dirac-fermion Hubbard model}
\label{sec:S2}

    \subsection{Sign problem behaviors}
        
        \begin{figure}[htbp]
            \centering
            \includegraphics{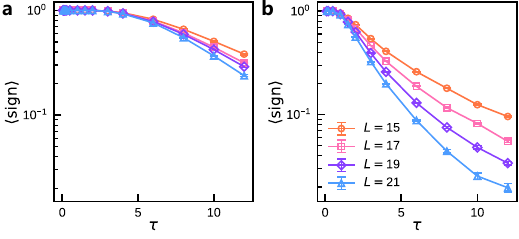}
            \caption{\textbf{Evolution of the average sign at the critical point.} \textbf{a}, Evolution from the DSM initial state. \textbf{b}, Evolution from the FM initial state.}
            \label{fig:FIG_SLAC_sign_relaxation}
        \end{figure}

        Fig.~\ref{fig:FIG_SLAC_sign_relaxation} shows that near the critical point of the single-Dirac-fermion Hubbard model, the average sign $\braket{\mathrm{sign}}$ decays in the imaginary-time relaxation with the Dirac semimetal (DSM) and the ferromagnetic (FM) initial states. We find that different initial states can have different decay rates. In the relaxation with the DSM initial state, the average sign decays more slowly. The data presented in Fig.~\ref{fig:SLAC} of the main text show the relaxation results starting from the DSM initial state. 

    \subsection{Verification of the QCP and critical exponents}
    \label{Sec:S2-B}

        Several physical quantities are needed to describe the universal scaling behaviors near the QCP. The spin structure factor $S({\bm{k}})$ with momentum $\bm{k}$ is defined as
        \begin{equation}
            S({\bm{k}}) \equiv \frac{1}{L^{2d}} \sum_{i,j} \mathrm{e}^{\mathrm{i} \bm{k}\cdot (\bm{r}_i-\bm{r}_j)} \langle{S_{i}^z S_{j}^z}\rangle,
        \end{equation}
        where the local spin operator is $S_{i}^{z} \equiv \bm{c}^\dagger_{i} \sigma^z \bm{c}_{i}$ with ${\bm{c}^\dagger\equiv({c_\uparrow^\dagger,c_\downarrow^\dagger}})$. 
        In addition, the correlation length ratio $R_{\rm FM}$ is defined as
        \begin{equation}
            R_{\rm FM} \equiv 1-\frac{S({\bm{k}}={\Delta \bm{k}})}{S ({\bm{k}}=0)},
            \label{eq:ratio}
        \end{equation}
        where $\Delta \bm{k} = \frac{1}{L} \bm{b_1} + \frac{1}{L} \bm{b_2}$ is the minimum momentum of electrons in a lattice with periodic boundary conditions, and $\bm{b_1}, \bm{b_2}$ are the reciprocal lattice vectors. 
        
        The fermion correlation function $G_{\rm f}(\bm{k})$ is defined as
        \begin{equation}
            G_{\rm f}(\bm{k})\equiv \frac{1}{L^2}\sum_{ij}\expup{\mathrm{i} \bm{k}(\bm{r}_i-\bm{r}_j) }\braket{c_{i\uparrow}^\dagger c_{j\downarrow}}.
            \label{eq:Gf_SLAC}
        \end{equation}
        We remark that the fermionic correlation $G_{\rm f}(\bm{k})$ is defined in momentum space, as commonly done in related works~\cite{Sorella2016prx,Lang2019prl,Vaezi2022prl,yu2023dirac,lang2025arxiv}. Compared to real-space correlations, which often decay in an oscillatory manner and thus exhibit somewhat subtle finite-size effects (e.g., as it may depend on the even-odd parity of the distance or the size), momentum-space correlations are often more straightforward and robust for analysis.
        Physically, the momentum-space fermionic correlation defined in Eqs.~\eqref{eq:Gf_SLAC}, \eqref{eq:Gf_tV} and \eqref{eq:Gf_SU3} captures the quasiparticle weight $Z$, which characterizes the discontinuity in the occupation number across the Fermi surface. Specifically, $Z = \lim_{L \to \infty} 2\left| G_{s\bar{s}}(\Delta \bm{k}) \right|$, where $s$ and $\bar{s}$ denote the two components of the Dirac spinor, corresponding to spin ($\uparrow,\downarrow$) or sublattice ($A,B$)  degrees of freedom in different models. In contrast, momentum-space correlations between the same spinor components are trivial, $G_{ss}(k) = 1/2$ for all $k$, due to spontaneous breaking of chiral symmetry. For detailed derivations and discussions, see Refs.~\cite{Sorella2016prx, yu2023dirac}.

        In the main text and figures of data, we abbreviate $S_{\rm FM}\equiv S(\bm{k}=\bm{0})$ and $G_{\rm f}\equiv G_{\rm f}(\bm{k}=\Delta\bm{k})$, unless otherwise specified.
        
        The correlation length ratio $R_{\rm FM}$ is a dimensionless quantity. In the nonequilibrium critical region, the correlation length ratio $R_{\rm FM}$ satisfies the following scaling form:
        \begin{equation}
            R_{\rm FM}(g,\tau,L) = f_R(gL^{1/\nu}, \tau L^{-z}),
            \label{eq:ratio_scaling}
        \end{equation}
        where $g = U - U_c$. To determine the quantum critical point $U_c$, we fix $\tau L^{-z}$ to be constant (e.g., we take $\tau L^{-z} = 0.3$ in the main text), so the scaling form of $R_{\rm FM}$ reduces to $R(g,\tau,L) = f_{R1}(gL^{1/\nu})$, which is similar to the traditional finite-size scaling. Accordingly, the critical point can be determined by the intersection of curves of $R_{\rm FM}$ versus $U$ for different $L$. We fit $U_c$ and $\nu$ based on the expansion of the scaling form:
        \begin{equation}
            R_{\rm FM}(g, L) = f_{R1}(gL^{1/\nu}) = \sum_{n=0}^{n_{\text{max}}} a_n g^n L^{n/\nu}.
            \label{eq:ratio_fitting}
        \end{equation}
        where we appropriately truncate the scaling functions with polynomials~\cite{Lang2019prl}.
        
\begin{figure}[htbp]
            \centering
            \includegraphics{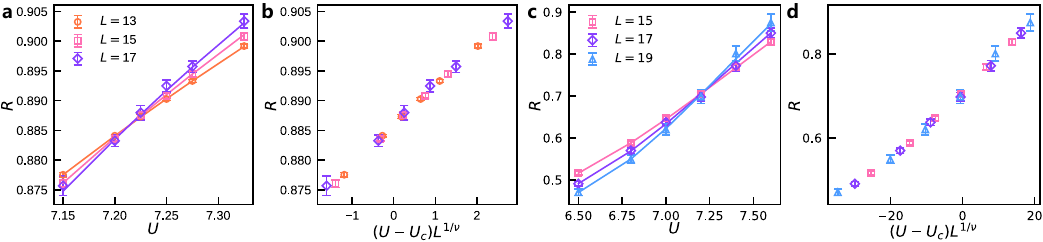}
            \caption{\textbf{Verification of quantum criticality using different initial states and imaginary times.} \textbf{a}--\textbf{b}, With the FM initial state and $\tau=0.5L^z$, the data collapse yields $U_c=7.214(44)$ and $\nu^{-1}=1.05(10)$. \textbf{c}--\textbf{d}, With the DSM initial state and $\tau=0.5L^z$, the data collapse yields $U_c=7.225(34)$ and $\nu^{-1}=1.16(9)$.}
            \label{fig:FIG_SLAC_Uc_sm}
        \end{figure}

        In the main text, based on the results from the DSM initial state, we fit according to Eq.~(\ref{eq:ratio_fitting}) and obtain $U_c=7.220(37), 1/\nu=1.18(3)$ at $\tau =0.3L^z$. Here we supplement data in Figs.~\ref{fig:FIG_SLAC_Uc_sm}\subref{c}-\ref{fig:FIG_SLAC_Uc_sm}\subref{d}, where $\tau$ is extended to $0.5L^{z}$, giving $U_c=7.225(34), 1/\nu=1.16(9)$ at $\tau =0.5L^z$, which is consistent with the results at $\tau =0.3L^z$ within error bars. This demonstrates that the imaginary time we used has already entered the nonequilibrium scaling regime and that the critical point values have converged. 
        On the other hand, we have also considered the FM initial state, as shown in Figs.~\ref{fig:FIG_SLAC_Uc_sm}\subref{a}-\ref{fig:FIG_SLAC_Uc_sm}\subref{b}, which gives $U_c=7.214(44), 1/\nu=1.05(10)$ at $\tau =0.5L^z$, also close to the results obtained from the DSM initial state.
        These results not only confirm the values of the critical point $U_c$ and the critical exponent $1/\nu$, but also show that the initial states can be chosen flexibly in our method, which provides a route to achieve reliable values of critical exponents by benchmarking the results with different initial states. This flexibility offers more possible options for alleviating the sign problem.

        The comparison of results from different initial states, as shown in Table~\ref{tab:SLAC_comparison}, provides an excellent bootstrap-style self-consistency check for the accuracy of our method, which we further illustrate below through the determination of the anomalous dimensions.

        \begin{table}[bt]
            \centering
            \caption{\textbf{Comparison of critical properties for the single-Dirac-fermion Hubbard model calculated by different methods.} The method used in this work is the nonequilibrium short-time PQMC.}
            \vspace{1em}
            \begin{tabularx}{\textwidth}{l *{4}{>{\centering\arraybackslash}X}}
               \toprule
                \textbf{Methods} & $\bm{U_c}$ & $\bm{\nu^{-1}}$ & $\bm{\eta_\phi}$ & $\bm{\eta_\psi}$ \\
                \midrule
                This work (from DSM, $\tau=0.3L^z$) & 7.220(37) & 1.18(3) & 0.36(3) & 0.134(3) \\
                This work (from DSM, $\tau=0.5L^z$) & 7.225(34) & 1.16(9) & - & - \\
                This work (from FM, $\tau=0.5L^z$) & 7.214(44) & 1.05(10) & 0.35(3) & 0.136(14) \\
                Gutzwiller-PQMC (equilibrium) \cite{Vaezi2022prl} & 7.275(25) & 1.19(3) & 0.31(1) & 0.136(5) \\
                FRG \cite{Vacca2015PRD} & - & 1.229 & 0.372 & 0.131 \\
                \bottomrule
            \end{tabularx}
            \label{tab:SLAC_comparison}
        \end{table}
        
\begin{figure}[htbp]
            \centering
            \includegraphics{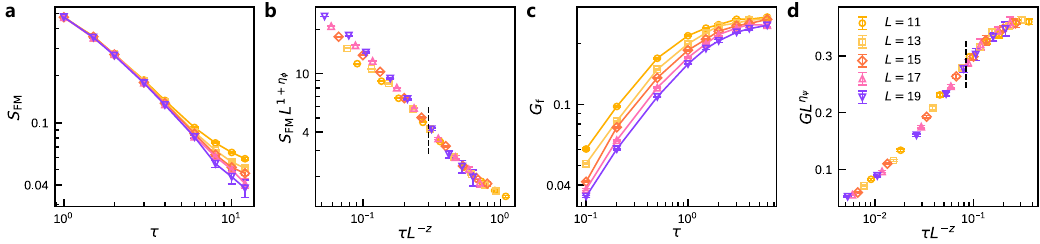}
            \caption{
                \textbf{Relaxation dynamics at QCP with FM initial state in single-Dirac-fermion Hubbard model.} 
                \textbf{a}-\textbf{b}, Curves of $S_{\rm FM}\equiv S(\bm{k}=\bm{0})$ versus $\tau$ for different sizes before and after rescaling.  
                \textbf{c}-\textbf{d}, Curves of $G_{\rm f}\equiv G_{\rm f}(\bm{k}=\Delta\bm{k})$ versus $\tau$ before and after rescaling.  Data collapse in the relaxation dynamics shows $\eta_\phi = 0.34(5)$ and $\eta_\psi = 0.131(20)$.
                Only the interval with sufficiently large $\tau L^{-z}$ is used for the data collapse, with the technical details shown in Fig.~\ref{fig:SLAC_datacollapse}.
            }
            \label{fig:SLAC_relaxation_FM}
        \end{figure}

        The anomalous dimensions of the bosonic field $\eta_\phi$ and the fermionic field $\eta_\psi$ can be determined by the following scaling relations:
        \equa{S(\bm{k}=\bm{0})=L^{-(1+\eta_\phi)}f_{S}\kuohao{gL^{1/\nu},\tau L^{-z}},\label{eq:SLAC_S_scaling}}
        \equa{G_{\rm f}(\bm{k}=\Delta\bm{k})=L^{-\eta_\psi}f_{G}\kuohao{gL^{1/\nu},\tau L^{-z}}.\label{eq:SLAC_G_scaling}}
        We obtain $\eta_\phi = 0.34(5)$ and $\eta_\psi = 0.131(20)$ from the data collapse in the relaxation dynamics at $g=0$, as shown in Fig.~\ref{fig:SLAC_relaxation_FM}.
        Note that only the data within the nonequilibrium scaling region are included in the data collapse analysis. Below we present the general principles and technical details for extracting the critical exponents.
    
\begin{figure}[bt]
            \centering
            \includegraphics[width=\linewidth]{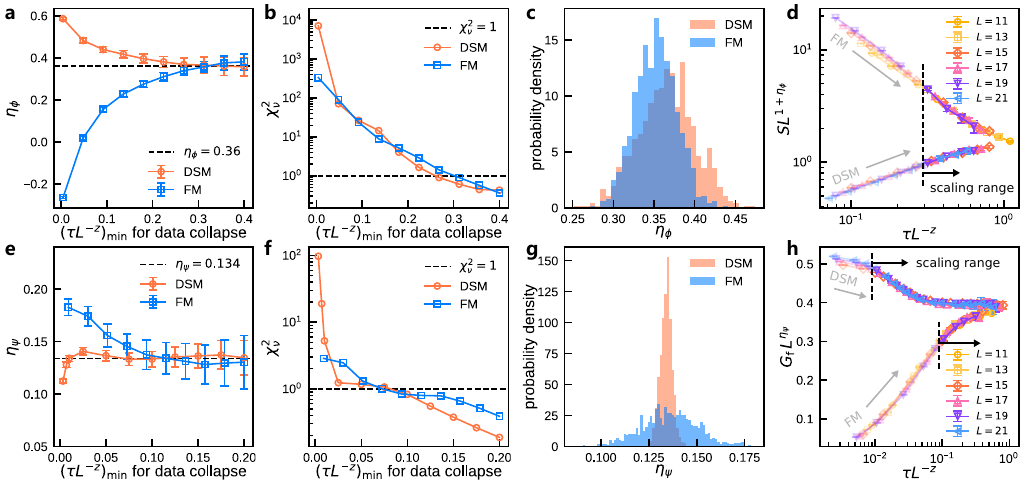}
            \caption{
                \textbf{Technical details of determining the critical exponents in the single-Dirac-fermion Hubbard model via data collapse.}
                \textbf{a}--\textbf{d}, Determination of the critical exponent $\eta_\phi$.
                \textbf{a}, Fitted values of $\eta_\phi$ versus the lower bound $\left(\tau L^{-z}\right)_{\text{min}}$ used in the data collapse analysis. The legend ``DSM'' denotes data collapse performed with the Dirac semi-metal initial state (the lower set of curves in \textbf{d}); the legend ``FM'' denotes data collapse performed with the ferromagnetic initial state (the upper set of curves in \textbf{d}). 
                \textbf{b}, Reduced $\chi_{\nu}^2$ of the data collapse versus  $\left(\tau L^{-z}\right)_{\text{min}}$. 
                \textbf{c}, Distribution of the fitted $\eta_\phi$ obtained from 1000 resamplings with $\left(\tau L^{-z}\right)_{\text{min}}=0.3$. The results are $\eta_\phi=0.36(3)$ with the DSM initial state and $\eta_\phi=0.35(3)$ with the FM initial state. 
                \textbf{d}, Scaling collapse of curves of the structure factor $S_{\rm FM}$ versus rescaled $\tau$ at $U_c$. 
                \textbf{e}--\textbf{h}, Determination of the critical exponent $\eta_\psi$.
                \textbf{e}, Fitted values of $\eta_\psi$ versus $\left(\tau L^{-z}\right)_{\text{min}}$. The legend ``DSM'' denotes data collapse performed with the DSM initial state (the upper set of curves in \textbf{h}); the legend ``FM'' denotes data collapse performed with the FM initial state (the lower set of curves in \textbf{h}). 
                \textbf{f}, Reduced $\chi_{\nu}^2$ of the data collapse versus $\left(\tau L^{-z}\right)_{\text{min}}$. 
                \textbf{g}, Distribution of the fitted $\eta_\psi$ obtained from 1000 resamplings with $\left(\tau L^{-z}\right)_{\text{min}}=0.01$ for the DSM initial state and $\left(\tau L^{-z}\right)_{\text{min}}=0.1$ for the FM initial state. The results are $\eta_\psi=0.134(3)$ for the DSM initial state and $\eta_\psi=0.136(14)$ for the FM initial state. 
                \textbf{h}, Scaling collapse of curves of the fermion correlation $G_{\rm f}$ versus rescaled $\tau$ at $U_c$. 
            }
            \label{fig:SLAC_datacollapse}
        \end{figure}

        As we stated in Sec.~\ref{sec:S1-E}, when $\tau L^{-z}$ is sufficiently large, the system enters the nonequilibrium scaling region, and the critical exponents can be accurately extracted during the nonequilibrium process without dependence on the initial state. In contrast, for too small values of $\tau L^{-z}$, the system remains too close to the initial state and lies outside the nonequilibrium scaling region governed by the critical point, where the scaling relations no longer hold.
        In practice, we apply this criterion to precisely determine the critical exponents and delineate the non-equilibrium scaling region.  During data collapse analysis, we systematically vary the lower bound of $\tau L^{-z}$ , denoted as $\left(\tau L^{-z}\right)_{\text{min}}$, and examine how the fitted critical exponents vary with $\left(\tau L^{-z}\right)_{\text{min}}$, as shown in Figs.~\ref{fig:SLAC_datacollapse}\subref{a} and \subref{e}. As $\left(\tau L^{-z}\right)_{\text{min}}$ for the data collapse increases, the fitted critical exponents gradually converge and, within the resolution of the error bars, no longer change with $\left(\tau L^{-z}\right)_{\text{min}}$. Moreover, the critical exponents obtained from nonequilibrium processes starting from different initial states converge to the same results within the error bars---this is the hallmark of having entered the nonequilibrium scaling region.

        Regarding the critical exponent $\eta_\phi$ in the single-Dirac-fermion Hubbard model, Fig.~\ref{fig:SLAC_datacollapse}\subref{a} shows that $\eta_\phi$ converges once $\left(\tau L^{-z}\right)_{\text{min}}>0.25$, and the results obtained from nonequilibrium processes starting from different initial states are consistent.  We use the reduced $\chi_{\nu}^2$ to assess the quality of the data collapse:
        \begin{equation}
            \chi_{\nu}^2 = \frac{1}{\nu}\sum_{i=1}^{N}\sum_L^{N_L}\frac{\kuohao{y_{iL}-\mu_i}^2}{\Delta y_{iL}^2} \, .
        \end{equation}
        For the rescaled curves corresponding to different system sizes $L$ ($N_L$ curves in total), we perform linear interpolation and then uniformly sample $N=50$ values of $\tau L^{-z}$ to obtain the curve ordinates $y_{iL}$ and their uncertainties $\Delta y_{iL}$. We then compute the weighted mean $\mu_i = \sum_L w_{iL}\, y_{iL} / \sum_L w_{iL}$ with weights $w_{iL}=1/\Delta y_{iL}^2$. The degrees of freedom for the reduced chi-square are $\nu = N (N_L - 1)$.
        As shown in Fig.~\ref{fig:SLAC_datacollapse}\subref{b}, when $\left(\tau L^{-z}\right)_{\text{min}}$ is very small, $\chi_{\nu}^2 \gg 1$, indicating very poor collapse quality and that the scaling form cannot describe such short $\left(\tau L^{-z}\right)_{\text{min}}$. As $\left(\tau L^{-z}\right)_{\text{min}}$ increases, $\chi_{\nu}^2$ decreases. Around $0.2<\left(\tau L^{-z}\right)_{\text{min}}<0.35$, $\chi_{\nu}^2$ approaches $1$, where the quality of the data collapse is optimal. Further increasing $\left(\tau L^{-z}\right)_{\text{min}}$ leads to $\chi_{\nu}^2<1$, which implies overfitting. This occurs because data points at larger $\tau L^{-z}$ suffer from more severe sign problems and thus have larger errors, exceeding the resolution of the collapse. 
        Taking into account the convergence behavior in Fig.~\ref{fig:SLAC_datacollapse}\subref{a}, the consistency between results from different initial states, and the collapse quality shown in Fig.~\ref{fig:SLAC_datacollapse}\subref{b}, we finally choose $\left(\tau L^{-z}\right)_{\text{min}}=0.3$ as the lower bound for $\tau L^{-z}$ in the data collapse analysis of the structure factor $S$ for the single-Dirac-fermion Hubbard model. We evaluate the uncertainty of the critical exponent using a resampling technique. As shown in Fig.~\ref{fig:SLAC_datacollapse}\subref{c}, we randomly perturb the data of the structure factor $S$ according to the size of the error bars of the original data and then perform the data collapse analysis again to extract $\eta_\phi$. Figure~\ref{fig:SLAC_datacollapse}\subref{c} presents the histogram of $\eta_\phi$ obtained from 1000 resamplings, showing consistent results between the DSM and FM initial states. By fitting the histogram with a Gaussian distribution, we obtain the critical exponent as $\eta_\phi=0.36(3)$ for the DSM initial state and $\eta_\phi=0.35(3)$ for the FM initial state. 
        Figure~\ref{fig:SLAC_datacollapse}\subref{d} displays the rescaled data collapse using these results. The black dashed line marks $\tau L^{-z}=0.3$, with the region to the right included in the scaling analysis, where curves of different system sizes collapse perfectly. The region to the left of the dashed line is outside the scaling regime, and the deviations from scaling can be seen from the degree of non-overlap between curves. Figure~\ref{fig:SLAC_datacollapse}\subref{d} also shows that the system has not yet evolved to equilibrium (which typically requires $\tau L^{-z}\sim 2$--$3$). Nevertheless, we are able to determine the accurate ground-state critical exponents from nonequilibrium data. Comparing Figs.~\ref{fig:SLAC_datacollapse}\subref{a} and \ref{fig:SLAC_datacollapse}\subref{d}, for the system sizes we studied, the $\tau L^{-z}$ needed for the convergence of $\eta_\phi$ is about one order of magnitude smaller than that required for the convergence of the structure factor $S$. At larger system sizes, the nonequilibrium critical region will be even broader.

        The same applies to other similar data collapse figures: our data collapse analysis is \textit{only} performed using data within the defined scaling range, which appears after a microscopic nonuniversal time scale. Data points at very small $\tau L^{-z}$ lie outside this range and are, therefore, excluded from the analysis. 
        However, it should be emphasized that the this is not the specific issue for the short-time scaling. Even for the more popular finite-size scaling, when the lattice size is too small to enter the scaling region, the finite-size scaling can lose its efficacy. 
        The boundary of this scaling range is critical for the practical application of our method, as it dictates the minimum permissible $\tau L^{-z}$ and thus the extent to which the sign problem can be alleviated. For this reason, we deliberately presented the crossover from scaling violation to scaling satisfaction in these figures. This approach provides a direct visualization of the scaling-range boundary, elucidating our rationale for selecting appropriate $\tau L^{-z}$ values and preventing the misconception that $\tau L^{-z}$ can be chosen arbitrarily small. We explicitly mark the scaling range in these figures with dash lines.

        From the above analysis procedure, it is clear that determining critical points and critical exponents using nonequilibrium scaling is highly controllable in terms of accuracy. We can assess the accuracy not only by examining the asymptotic convergence and the quality of the data collapse, but more importantly, by the fact that nonequilibrium evolutions from different initial states are governed by the same ground-state critical exponents. The consistency of the critical points and critical exponents obtained with different initial states is smoking gun evidence that our results are sufficiently accurate and self-consistent. This is a significant advantage of the nonequilibrium approach compared with finite-temperature scaling or finite-size scaling in the conventional equilibrium approaches. Even in situations where the sign problem is particularly severe and prevents accurate results within the scaling range, the early-time results from different initial states can still bracket a controlled range for the critical points and critical exponent.

        Table~\ref{tab:SLAC_comparison} compares the results in the present work with those of other methods. The consistent results shown in this table demonstrate that the method based on the short-time dynamics can accurately determine the critical properties of the ground-state QCP with much fewer computational costs.

    \subsection{Procedure for determining the dynamical exponent \nota{z}}
        
        The quantum phase transitions studied in this work fall into the Dirac QCP universality class, for which the dynamical exponent is known to be $z=1$. However, applying this method to systems with an unknown dynamical exponent requires a self-consistent treatment, as both $z$ and the critical point should be determined together. To address this, we introduce a general procedure for extracting $z$ without prior knowledge of the critical point. We then demonstrate the efficacy of this procedure by applying it to the single-Dirac-fermion Hubbard model, providing a detailed analysis of the numerical results.

        \begin{figure}[htbp]
            \centering
            \includegraphics{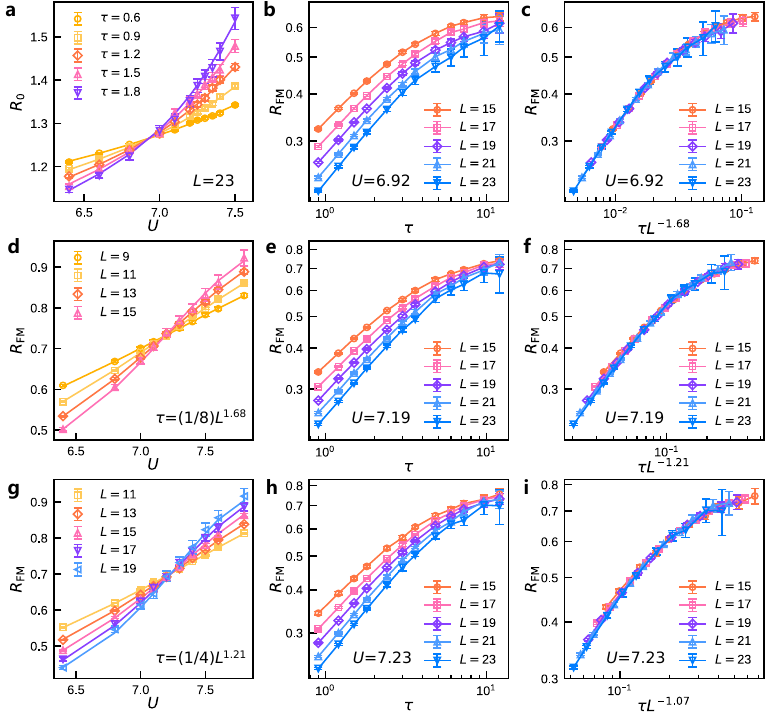}
            \caption{
                \textbf{Determining the critical point $U_c$ and the dynamical exponent $z$ for the single-Dirac-fermion Hubbard model.} 
                \textbf{a}, Curves of dimensionless ratio $R_0(U,\tau) \equiv S_{\text{FM}}(U,2\tau)/S_{\text{FM}}(U,\tau)$ versus $U$ at different $\tau$. The crossing point gives an initial estimate of the critical point $U_c = 6.92(1)$. 
                \textbf{b}, Relaxation dynamics of the correlation length ratio $R_{\text{FM}}$ at $U = 6.92$. 
                \textbf{c}, Data collapse of \textbf{b} yields an initial estimate of the dynamical exponent $z = 1.68(7)$. 
                \textbf{d}, By fixing $\tau = \frac{1}{8} L^{1.68}$, the crossing point of $R_{\text{FM}}$ versus $U$ curves gives an improved estimate of the critical point $U_c = 7.19(1)$. 
                \textbf{e}, Relaxation dynamics of the correlation length ratio $R_{\text{FM}}$ at $U = 7.19$. 
                \textbf{f}, Data collapse of \textbf{e} yields an improved estimate of the dynamical exponent $z = 1.21(14)$.
                \textbf{g}, By fixing $\tau = \frac{1}{4} L^{1.21}$, the crossing point of $R_{\text{FM}}$ versus $U$ curves gives an improved estimate of the critical point $U_c = 7.23(1)$. 
                \textbf{h}, Relaxation dynamics of the correlation length ratio $R_{\text{FM}}$ at $U = 7.23$. 
                \textbf{i}, Data collapse of \textbf{e} yields an improved estimate of the dynamical exponent $z = 1.07(9)$.
            }
            \label{fig:SLAC_z}
        \end{figure}
        
        We introduce a new dimensionless ratio:
        \begin{equation}
            R_0(U,\tau) \equiv \frac{S_{\text{FM}}(U,2\tau)}{S_{\text{FM}}(U,\tau)},
        \end{equation}
        where $S_{\text{FM}}$ is the zero-momentum structure factor. Suppose we do not know the precise values of $U_c$ and $z$ initially; we determine them via the following iterative steps:
        \begin{enumerate}[label={}]
            \item \textbf{Step 1. }
            We first select a relatively large system size $L$, for instance, $L=23$ in Fig.~\ref{fig:SLAC_z}\subref{a}, such that the factor $\tau L^{-z}$ has negligible influence on the scaling of $R_0$. According to the scaling form $R_0=2^{\frac{1+\eta_\phi}{z}}f_{R_0}(g\tau^{\frac{1}{\nu z}})$, the crossing point of $R_0$ curves for different $\tau$ provides an initial estimate of the critical point, $U_c = 6.92(1)$, as shown in Fig.~\ref{fig:SLAC_z}\subref{a}. Note that this estimate of $U_c$ may not yet be highly accurate, since the finite-size effects involving $\tau L^{-z}$ have only been approximately neglected.
            \item \textbf{Step 2. }
            At the tentative critical point $U=6.92$ obtained from Step 1, we estimate the dynamical exponent $z$ from the scaling form of the correlation length ratio, $R_{\text{FM}}=f_{R}(\tau L^{-z})$. In Fig.~\ref{fig:SLAC_z}\subref{b}-\ref{fig:SLAC_z}\subref{c}, by performing data collapse of the relaxation dynamics of $R_{\text{FM}}$, we obtain an initial estimate for the dynamical exponent: $z=1.68(7)$. Again, this initial estimate may still have uncertainties.
            \item \textbf{Step 3. }
            Next, we fix the value of $\tau L^{-z}$ using the estimate $z=1.68$ from Step 2 and refine the critical point determination using the nonequilibrium method described in the main text. As shown in Fig.~\ref{fig:SLAC_z}\subref{d}, we obtain an improved estimate of the critical point, $U_c=7.19(1)$.
            \item \textbf{Further iterations.} 
            Returning to Step 2, we re-estimate $z$ at the updated critical point $U=7.19$. As shown in Fig.~\ref{fig:SLAC_z}\subref{e}-\ref{fig:SLAC_z}\subref{f}, this yields an improved value $z=1.21(14)$. One can further iterate Steps 2 and 3 to progressively enhance the accuracy of both $U_c$ and $z$.
        \end{enumerate}
        By repeatedly iterating Steps 2 and 3, the estimates for both $U_c$ and $z$ converge towards stable values. In Fig.~\ref{fig:SLAC_z}, after three rounds of iterations, we obtain final estimates of $U_c=7.23(1)$ and $z=1.07(9)$, which are consistent within error bars with the result $U_c=7.220(37)$ presented in the main text and the known Dirac QCP value of $z=1$.
        
        This procedure can also be generally applied to other models and different types of QCP, even when both the location of the critical point and the value of $z$ are initially unknown.

\section{More details for the spinless \nota{t}-\nota{V} model}
\label{sec:S3}

    \subsection{Sign problem behaviors}

\begin{figure}[htbp]
            \centering
            \includegraphics{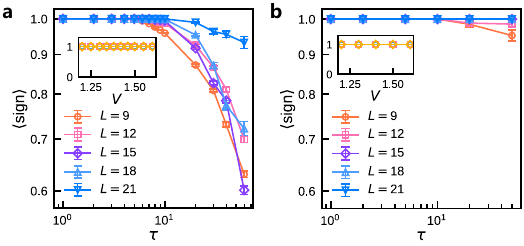}
            \caption{\textbf{Evolution of average sign at the QCP (inset shows $\braket{\mathrm{sign}}$ versus $V$ near the QCP for $\tau=0.3L^z$).} \textbf{a}, Evolution from the DSM initial state. \textbf{b}, Evolution from the CDW initial state.}
            \label{fig:tV_sign}
        \end{figure}

        As shown in Fig.~\ref{fig:tV_sign}, for both DSM initial state and CDW initial state, the average sign in the short-time stage is close to one, indicating the sign problem is very weak. In particular, for $\tau=0.3L^z$, as shown in the insets of Fig.~\ref{fig:tV_sign}, the system almost remains sign-free near the critical point $V_c=1.35(1)$ determined in the main text.

        Incidentally, Fig.~\ref{fig:tV_sign} shows that the decay rate of $\braket{\mathrm{sign}}$ does not monotonically change with the system size $L$. In some cases, a larger system size $L$ results in a weaker sign problem. Such phenomena have also been reported in the equilibrium projector QMC study~\cite{li2022asymptotic} and finite-temperature DQMC~\cite{Scalettar2015prb}.

    \subsection{Verification of the QCP and critical exponents}

     For this model, the ordered phase is the charge density wave (CDW) state, for which the structure factor $S_{\rm CDW}$ can be defined as:
        \begin{equation}
            S_{\rm CDW}({\bm{k}}) \equiv \frac{1}{L^{2d}} \sum_{i,j} \mathrm{e}^{\mathrm{i} \bm{k}\cdot (\bm{r}_i-\bm{r}_j)} \langle{m_{i} m_{j}}\rangle,
        \end{equation}
        where the local order operator is $m_{i} \equiv \frac{1}{2} (n_{i,A}-n_{i,B})$, which represents the difference in particle number density between the two sublattices in a unit cell. 
        The associated correlation length ratio $R_{\rm CDW}$ is defined as:
        \begin{equation}
            R_{\rm CDW} \equiv 1-\frac{S_{\rm CDW}({\bm{k}}={\Delta \bm{k}})}{S_{\rm CDW}(\bm{k}=0)}.
        \end{equation}
        The fermion correlation $G_{\rm f}(\bm{k})$ is defined similarly to Eq.~\eqref{eq:Gf_SLAC}:
        \begin{equation}
            G_{\rm f}(\bm{k})\equiv \frac{1}{L^2}\sum_{ij}\expup{\mathrm{i} \bm{k}(\bm{r}_i-\bm{r}_j) }\braket{c_{i,A}^\dagger c_{j,B}}.
            \label{eq:Gf_tV}
        \end{equation}
        Here, the sublattice indices $A$ and $B$ function as Dirac spinor indices.
        In the main text and figures of data, we abbreviate $S_{\rm CDW}\equiv S_{\rm CDW}(\bm{k}=\bm{0})$ and $G_{\rm f}\equiv G_{\rm f}(\bm{k}=\mathrm{K}+\Delta\bm{k})$, where $\mathrm{K}=\left(\pm\frac{4\pi}{3}, 0\right)$ represents the momentum at the Dirac points.
        
        \begin{figure}[htbp]
            \centering
            \includegraphics{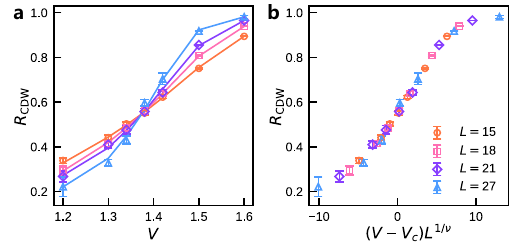}
            \caption{
                \textbf{Determination of the critical point and $1/\nu$ with DSM initial state.} \textbf{a}, At $\tau = 0.3L^z$, curves of $R_{\rm CDW}$ versus $V$ for different sizes intersect at $V_c = 1.37(2)$. \textbf{b}, Data collapse of curves of $R_{\rm CDW}$ versus rescaled $(V-V_c)$ with $\nu = 0.79(5)$.
            }
            \label{fig:tV_Vc_DSM}
        \end{figure}

        \begin{figure}[htbp]
            \centering
            \includegraphics{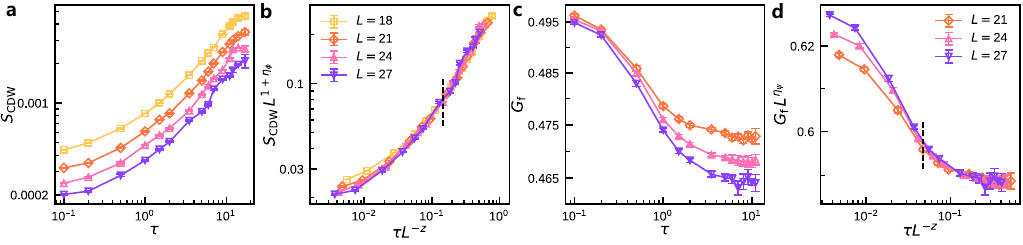}
            \caption{
                \textbf{Relaxation dynamics at QCP with DSM initial state in spinless $t$-$V$ model.} 
                \textbf{a}-\textbf{b}, Curves of $S_{\rm CDW}\equiv S_{\rm CDW}(\bm{k}=\bm{0})$ versus $\tau$ for different sizes before and after rescaling.  
                \textbf{c}-\textbf{d}, Curves of $G_{\rm f}\equiv G_{\rm f}(\bm{k}=K+\Delta\bm{k})$ versus $\tau$ before and after rescaling.  Data collapse in the relaxation dynamics shows $\eta_\phi = 0.44(2)$ and $\eta_\psi = 0.072(4)$.
            }
            \label{fig:tV_relaxation_DSM}
        \end{figure}

        In Fig.~\ref{fig:tV} of the main text, we showed the results of determination of critical properties via the short-time dynamics from the ordered CDW initial state. In contrast, here we show consistent results can also be obtained from the DSM initial state.
        
        Fig.~\ref{fig:tV_Vc_DSM} illustrates the correlation length ratio $R_{\rm CDW}$ as a function of interaction strength $V$ at $\tau = 0.3L^z$. We fit the data in Fig.~\ref{fig:tV_Vc_DSM}\subref{a} using the scaling form in Eq.~\ref{eq:ratio_fitting}, determining the intersection point of the curves for different sizes as $V_c = 1.37(2)$ and the scaling collapse exponent as $\nu = 0.79(5)$. These results are close to those obtained by equilibrium methods~\cite{Wang2014njp,li2015njp} and the short-time dynamics with ordered CDW initial state discussed in the main text.
        Fig.~\ref{fig:tV_relaxation_DSM} shows the relaxation dynamics of $S_{\rm CDW}= S_{\rm CDW}(\bm{k}=0)$ and $G_{\rm f}= G_{\rm f}(\bm{k}=\mathrm{K}+\Delta \bm{k})$ starting from the DSM initial state. Their scaling behavior is described by Eq.~(\ref{eq:SLAC_S_scaling}) and Eq.~(\ref{eq:SLAC_G_scaling}). According to the scaling collapse of the evolution of $S_{\rm CDW}$ in Fig.~\ref{fig:tV_relaxation_DSM}\subref{a}-\ref{fig:tV_relaxation_DSM}\subref{b}, one finds $\eta_\phi = 0.44(2)$. In addition, from the scaling collapse of the evolution of $G_{\rm f}$ in Fig.~\ref{fig:tV_relaxation_DSM}\subref{c}-\ref{fig:tV_relaxation_DSM}\subref{d}, we obtain $\eta_\psi = 0.072(4)$. The value of $\eta_\phi$ is close to that obtained by the equilibrium methods~\cite{Wang2014njp,li2015njp} and the short-time dynamics with ordered CDW initial state discussed in the main text; while the value of $\eta_\psi$ is close to the FRG result~\cite{Vacca2015PRD}, and also consistent with that obtained from the CDW initial state shown in the main text. Consequently, the results of critical properties in the $t$-$V$ model further demonstrate the efficiency and accuracy of our method. More crucially, for the first time, through our method we achieve the reliable result of fermionic anomalous dimension $\eta_\psi$ for the QCP in the spinless $t$-$V$ model by unbiased QMC simulation, as shown in Table~\ref{tab:tV_comparison}.

        \begin{table}[htbp]
            \centering
            \caption{\textbf{Comparison of critical properties for the spinless $t$-$V$ model calculated by different methods.} The method used in this work is the nonequilibrium short-time PQMC.}
            \vspace{1em}
            \begin{tabularx}{\textwidth}{l *{4}{>{\centering\arraybackslash}X}}
               \toprule
                \textbf{Methods} & $\bm{U_c}$ & $\bm{\nu}$ & $\bm{\eta_\phi}$ & $\bm{\eta_\psi}$ \\
                \midrule
                This work (from CDW, $\tau=0.3L^z$) & 1.35(1) & 0.77(12) & 0.49(5) & 0.073(4) \\
                This work (from DSM, $\tau=0.3L^z$) & 1.37(2) & 0.79(5) & 0.44(2) & 0.072(4) \\
                Majorana QMC (equilibrium)~\cite{li2015njp} & 1.355(1) & 0.77(2) & 0.45(2) & - \\
                Continuous-time QMC (equilibrium)~\cite{Wang2014njp} & 1.356(1) & 0.80(3) & 0.302(7) & - \\
                FRG~\cite{Vacca2015PRD} & - & 0.929 & 0.602 & 0.069 \\
                \bottomrule
            \end{tabularx}
            \label{tab:tV_comparison}
        \end{table}

\section{More details for the SU(3) Hubbard model}
\label{sec:S4}

    \subsection{SU(3) algebra}
    
        The $\rm SU(3)$ Hubbard model with staggered flux is invariant under $\rm SU(3)$ transformations in the flavor space of fermions. Here we give a brief review on the $\rm SU(3)$ algebra.

        The generators of the $\rm SU(3)$ group are represented by the well-known Gell-Mann matrices. Below are the eight Gell-Mann matrices, corresponding to the generators of $\rm SU(3)$ group:
        \begin{equation}
            \begin{aligned}
                \lambda_1 = \begin{pmatrix} 
                0 & 1 & 0 \\ 
                1 & 0 & 0 \\ 
                0 & 0 & 0 
                \end{pmatrix},
                \quad
                \lambda_2 = \begin{pmatrix} 
                0 & -i & 0 \\ 
                i & 0 & 0 \\ 
                0 & 0 & 0 
                \end{pmatrix},
                \quad
                \lambda_3 = \begin{pmatrix} 
                1 & 0 & 0 \\ 
                0 & -1 & 0 \\ 
                0 & 0 & 0 
                \end{pmatrix},\\
                \lambda_4 = \begin{pmatrix} 
                0 & 0 & 1 \\ 
                0 & 0 & 0 \\ 
                1 & 0 & 0 
                \end{pmatrix},
                \quad
                \lambda_5 = \begin{pmatrix} 
                0 & 0 & -i \\ 
                0 & 0 & 0 \\ 
                i & 0 & 0 
                \end{pmatrix},
                \quad
                \lambda_6 = \begin{pmatrix} 
                0 & 0 & 0 \\ 
                0 & 0 & 1 \\ 
                0 & 1 & 0 
                \end{pmatrix},\\
                \lambda_7 = \begin{pmatrix} 
                0 & 0 & 0 \\ 
                0 & 0 & -i \\ 
                0 & i & 0 
                \end{pmatrix},
                \quad
                \lambda_8 = \begin{pmatrix} 
                1/\sqrt{3} & 0 & 0 \\ 
                0 & 1/\sqrt{3} & 0 \\ 
                0 & 0 & -2/\sqrt{3} 
                \end{pmatrix}.
            \end{aligned}
        \end{equation}
        Note that only $\lambda_8$ is full rank, the other seven are not. That is to say, only the order associated with $\lambda_8$ can gap out all flavors of fermions.

        The Lie algebra structure of the $\rm SU(3)$ group is determined by the commutation relations of its generators. The commutation relations between the generators of the $\rm SU(3)$ group are listed as follows:
        \begin{equation}
        [\lambda_a, \lambda_b] = 2\mathrm{i} \sum_c f_{abc} \lambda_c,
        \end{equation}
        where \(f_{abc}\) are the structure constants, specifically:
        \begin{equation}
        f_{123} = 1,
        \end{equation}
        \begin{equation}
        f_{147} = f_{246} = f_{257} = f_{345} = \frac{1}{2},
        \end{equation}
        \begin{equation}
        f_{156} = f_{367} = -\frac{1}{2},
        \end{equation}
        \begin{equation}
        f_{458} = f_{678} = \frac{\sqrt{3}}{2}.
        \end{equation}
        These commutation relations will determine the manifold shape of the ground-state degeneracy space of the ordered phase, which we will further analyze with numerical evidence in the subsequent sections.

    \subsection{Mean-field analyses}

        \begin{figure}[htbp]
            \centering
            \includegraphics{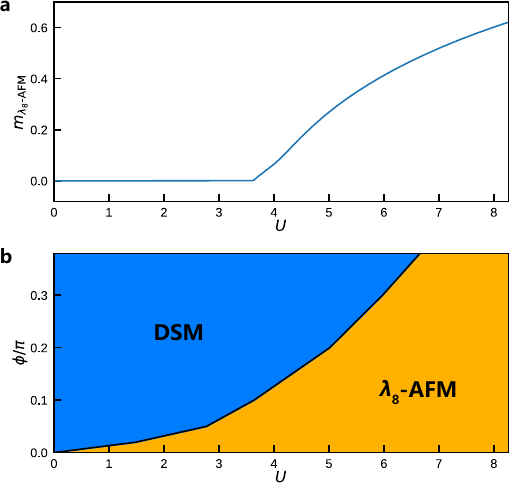}
            \caption{\textbf{Mean-field results for the $\lambda_8$-AFM order.} \textbf{a}, For $\phi=0.1\pi, L=42$, a DSM-AFM transition in the mean-field approximation is seen at $U=3.637(5)$. \textbf{b}, Mean-field phase diagram with $L=42$.}
            \label{fig:SU3_MeanField}
        \end{figure}

        To qualitatively understand the salient features of ground-state phase diagram and the dominant symmetry spontaneous breaking ordering in the SU(3) Hubbard model, we first perform a mean-field analysis. We rewrite the interaction term in the following form and apply the mean-field approximation:
        \begin{equation}
            \begin{aligned}
                \frac{U}{2} \sum_i \left(\sum_{\alpha} n_{i\alpha}-\frac{3}{2}\right)^2 
                =\, & \frac{U}{2} \sum_i \sum_{\alpha,\alpha'} n_{i\alpha} n_{i\alpha'} + \mathrm{const.} \\
                =\, & -\frac{U}{4} \sum_i \sum_{\alpha,\alpha'} \left(n_{i\alpha} - n_{i\alpha'}\right)^2 + \mathrm{const.} \\
                =\, & -\frac{3U}{16} \sum_i \sum_{n} \left(\bm{c}_i^\dagger \lambda_n \bm{c}_i \right)^2 + \mathrm{const.} \\
                \approx \, & -\frac{3U}{16} \sum_{n} \braket{m_{\lambda_n\text{-AFM}}}  \sum_i (-1)^i \bm{c}_i^\dagger \lambda_n \bm{c}_i + \mathrm{const.} \, ,
            \end{aligned}
        \end{equation}
        where $\bm{c}_i^\dagger\equiv\mat{\bm{c}_{i1}^\dagger&\bm{c}_{i2}^\dagger&\bm{c}_{i3}^\dagger}$, and $\lambda_n$=($\lambda_1, \lambda_2, \dots, \lambda_8$) are the eight Gell-Mann matrices. The mean-field order parameter is defined as
        \begin{equation}
            \braket{m_{\lambda_n\text{-AFM}}} \equiv \frac{1}{L^2} \sum_i (-1)^i \braket{\bm{c}_i^\dagger \lambda_n \bm{c}_i}
        \end{equation}
        After applying the mean-field approximation, the Hamiltonian is entirely expressed as a quadratic form of fermionic operators and can be solved for the ground state using exact diagonalization. The problem is then simply reduced to a self-consistent calculation of the order parameter $\left\{m_{\lambda_n\text{-AFM}}\right\}$. Fig.~\ref{fig:SU3_MeanField}\subref{a} shows the variation of the order parameter $m_{\lambda_8\text{-AFM}}$ with interaction strength $U$ calculated using the mean-field method. When the interaction is strong, $m_{\lambda_8\text{-AFM}}$ starts to increase with $U$, showing clear characteristics of a continuous phase transition. We control different magnetic flux $\phi$ to find the corresponding phase boundary $U$, as shown in the mean-field phase diagram in Fig.~\ref{fig:SU3_MeanField}\subref{b}.

        We can understand why the $\lambda_8$-AFM order is preferred by examining the ground state energy of the system. Consider placing the eight different order parameter operators in the same external field $h$:
        \begin{equation}
            H^{\text{sat}}_n = -h\sum_i (-1)^i \bm{c}_i^\dagger \lambda_n \bm{c}_i,
        \end{equation}
        whose ground state corresponds to the saturated $\lambda_n$-AFM ordered state. At half-filling, the ground state energy of $H^{\text{sat}}_8(h)$ is $-\frac{2}{\sqrt{3}}h L^2$. The energy gap at half-filling, between the ground state of $H^{\text{sat}}_8(h)$ and the first excited state of $H^{\text{sat}}_8(h)$, is a finite value $\frac{2}{\sqrt{3}}h$. For $H^{\text{sat}}_{n \ne 8}(h)$, the ground state energy is $-h L^2$ and the energy gap is $0$. 
        In other words, even considering the saturated ordered state, only the $\lambda_8$-AFM order can open a gap in Dirac fermions, while other types of $\lambda_n$-AFM cannot. 

        \begin{figure}[htbp]
            \centering
            \includegraphics{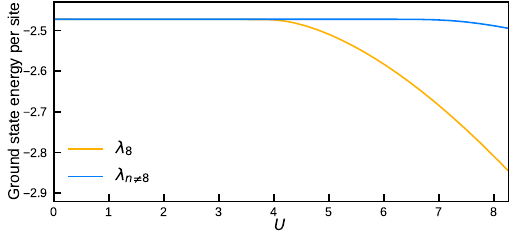}
            \caption{\textbf{Mean-field ground state energy per site at half-filling of different types of $\lambda_n$-AFM.} For the mean-field calculation with $\phi=0.1, L=42$, the ground state energy of the $\lambda_8$-AFM mean-field Hamiltonian is lower than that of other types of Hamiltonians.}
            \label{fig:SU3_MeanField_GSenergy}
        \end{figure}
        
        Next, we calculate the mean-field ground-state energy for different $\lambda_n$-AFM orders. From Fig.~\ref{fig:SU3_MeanField_GSenergy}, we find that $\lambda_8$-AFM mean-field states have lower ground-state energy than other types of antiferromagnetic orders in the whole interaction regime of the ordered phases. Thus, from a mean-field perspective, the system favors the $\lambda_8$-AFM order.

    \subsection{Phase boundary and critical exponents}

        Here, we perform QMC simulation through short-time relaxation to systematically study the phase boundary and critical properties of the quantum phase transition. To determine the phase boundary of the $\rm SU(3)$ Hubbard model with staggered flux, we compute the structure factor and correlation length ratio. For the $\lambda_8$-AFM order, the structure factor is defined as:
        \begin{equation}
            S_{\lambda_8\text{-AFM}}({\bm{k}}) = \frac{1}{L^{2d}} \sum_{i,j} \mathrm{e}^{\mathrm{i} \bm{k}\cdot (\bm{r}_i-\bm{r}_j)} (-1)^{i+j}\braket{ \bm{c}_i^\dagger \lambda_8 \bm{c}_i \bm{c}_j^\dagger \lambda_8 \bm{c}_j },
        \end{equation}
        The associated correlation length ratio is defined as 
        \begin{equation}
            R_{\lambda_8\text{-AFM}}\equiv 1-\frac{S_{\lambda_8\text{-AFM}}({\bm{k}}={\Delta \bm{k}})}{S_{\lambda_8\text{-AFM}}(\bm{k}=0)}.
        \end{equation}
        The fermion correlation $G_{\rm f}(\bm{k})$ is defined the same as Eq.~\eqref{eq:Gf_tV}:
        \begin{equation}
            G_{\rm f}(\bm{k})\equiv \frac{1}{L^2}\sum_{ij}\expup{\mathrm{i} \bm{k}(\bm{r}_i-\bm{r}_j) }\braket{c_{i,A}^\dagger c_{j,B}},
            \label{eq:Gf_SU3}
        \end{equation}
        where $A,B$ represent two inequivalent sublattices in a square lattice with staggered flux.
        In the main text and data figures, unless otherwise specified, we omit the momentum variables and denote $S_{\lambda_8\text{-AFM}}\equiv S_{\lambda_8\text{-AFM}}(\bm{k}=0)$ and $G_{\rm f}\equiv G_{\rm f}(\bm{k}=\mathrm{K}+\Delta\bm{k})$, where $\mathrm{K}=\left(\pm\frac{\pi}{2}, \pm\frac{\pi}{2}\right)$ represents the momentum at the Dirac points.

        \begin{figure}[htbp]
            \centering
            \includegraphics{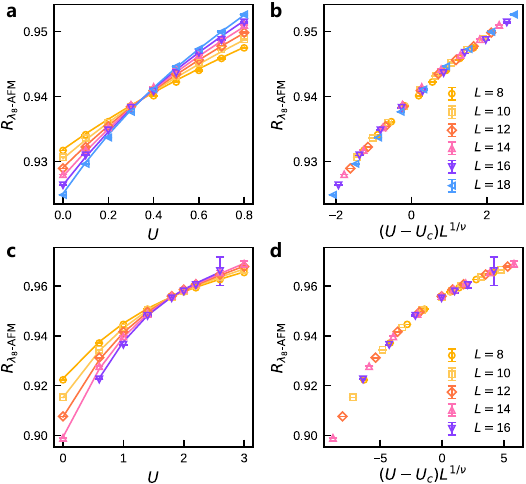}
            \caption{
                \textbf{Curves of $R_{\lambda_8\text{-AFM}}$ versus $U$ for different sizes with $\lambda_8$-AFM initial state at $\tau = 0.25L^z$.} 
                \textbf{a}-\textbf{b}, For $\phi=0.05\pi$, the fitted critical point is $U_c=0.347(4)$, with the critical exponent $1/\nu=0.67(1)$. 
                \textbf{c}-\textbf{d}, For $\phi=0.1\pi$, the fitted critical point is $U_c=2.0(3)$, with $1/\nu=0.58(7)$.
            }
            \label{fig:SU3_QCP}
        \end{figure}

        \begin{figure}[htbp]
            \centering
            \includegraphics{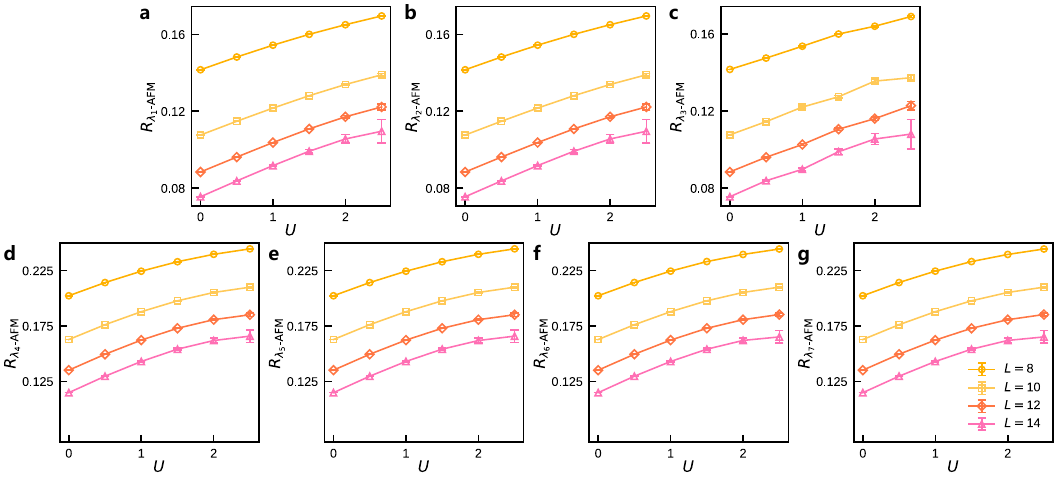}
            \caption{{\bf Correlation length ratios for $\lambda_1$-AFM to $\lambda_7$-AFM versus $U$ for different sizes with $\lambda_8$-AFM initial state at $\tau = 0.25L^z, \phi=0.075\pi$.} There is no crossing point in the curves of $R_{\lambda_n\text{-AFM}}$ versus $U$.}
            \label{fig:SU3_lambda1-7}
        \end{figure}
        
        To identify phase boundary between the DSM phase and the $\lambda_8$-AFM phase, we prepare a \nota{\lambda_8}-AFM initial state and calculate the critical point for fixed $\phi$ by the method of short-time dynamics. For fixed \nota{\tau L^{-z}=0.25}. We calculate the critical point for $\phi=0.05\pi$, $\phi=0.075\pi$ and $0.1\pi$ by the intersection points of curves of $R_{\lambda_8\text{-AFM}}$ versus $U$ for different $L$. As shown in Fig.~\ref{fig:SU3_QCP}, critical points are $U_c=0.347(4)$ for $\phi=0.05\pi$, $U_c=2.0(3)$ for $\phi=0.1\pi$, and $U_c=1.10(5)$ for $\phi=0.075\pi$ as shown in the main text.

        Here, to further confirm the \nota{\lambda_8}-AFM is the dominant ordering, we also compute the correlation length ratio for other quantities.  We find that there is no crossing in the curves of $R_{\lambda_n\text{-AFM}}$ for the AFM orders defined by $\lambda_1$ to $\lambda_7$ versus $U$, as shown in Fig.~\ref{fig:SU3_lambda1-7}. The results of the correlation length ratio decrease with $L$, demonstrating the absence of long-range order. Hence, the $\lambda_8$-AFM is the dominant ordering, and other AFM orders are all short-range in the half-filled $\rm SU(3)$ Hubbard model with staggered flux. 

        Moreover, scaling collapse for the curves of $R_{\lambda_8\text{-AFM}}$ versus rescaled $(U-U_c)$ gives the value of $1/\nu$. Accordingly, we obtain $1/\nu=0.67(1)$ and $1/\nu=0.58(7)$ for $\phi=0.05\pi$ and $\phi=0.1\pi$, respectively. Combining $1/\nu=0.68(5)$ with $\phi=0.075\pi$ shown in the main text, we find that the values of $1/\nu$ are consistent with each other within error bar, showing the universality of the quantum phase transition between DSM and \nota{\lambda_8}-AFM ordered phase.
              
\begin{figure}[htbp]
            \centering
            \includegraphics{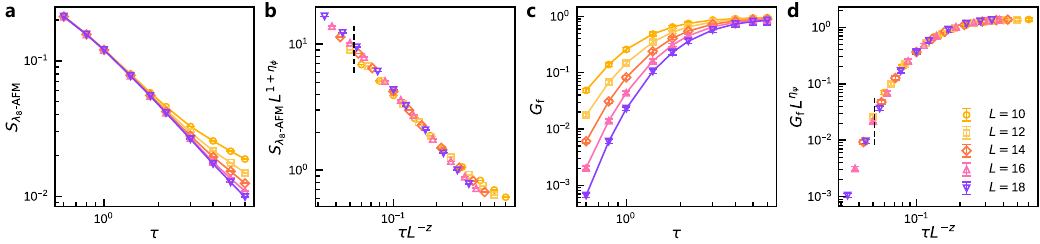}
            \caption{
                \textbf{Relaxation dynamics at the QCP $\phi=0.05\pi,U=0.347(4)$ with $\lambda_8$-AFM initial state in SU(3) Hubbard model.} 
                \textbf{a}-\textbf{b}, Curves of $S$ versus $\tau$ for different sizes before and after rescaling.  
                \textbf{c}-\textbf{d}, Curves of $G$ versus $\tau$ before and after rescaling.  Data collapse in the relaxation dynamics shows $\eta_\phi = 0.51(3)$ and $\eta_\psi = 0.16(2)$.
            }
            \label{fig:SU3_relaxation_005}
        \end{figure}

\begin{figure}[htbp]
            \centering
            \includegraphics{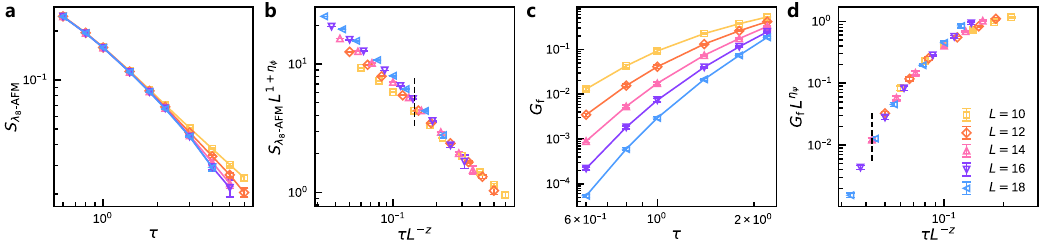}
            \caption{
                \textbf{Relaxation dynamics at the QCP $\phi=0.1\pi,U=2.0(3)$ with $\lambda_8$-AFM initial state in SU(3) Hubbard model.} 
                \textbf{a}-\textbf{b}, Curves of $S$ versus $\tau$ for different sizes before and after rescaling.  
                \textbf{c}-\textbf{d}, Curves of $G$ versus $\tau$ before and after rescaling.  Data collapse in the relaxation dynamics shows $\eta_\phi = 0.58(10)$ and $\eta_\psi = 0.16(3)$.
            }
            \label{fig:SU3_relaxation_01}
        \end{figure}

        To determine the anomalous dimensions $\eta_\phi$ and $\eta_\psi$ of the bosonic field and the fermionic field we study the relaxation dynamics of the structure factor $S_{\lambda_8\text{-AFM}}=S_{\lambda_8\text{-AFM}}(\bm{k}=0)$ and the fermion correlation $G_{\rm f}=G_{\rm f}(\bm{k}=\mathrm{K}+\Delta\bm{k})$ at the critical point. The curves before and after rescaling are shown in Figs.~\ref{fig:SU3_relaxation_005} and \ref{fig:SU3_relaxation_01} and Fig.~\ref{fig:SU3} in the main text for different $\phi$. The critical exponents determined above are summarized in Table~\ref{tab:SU3_critical}, from which we find that the anomalous dimensions for different $\phi$ are consistent with each other within error bar, further confirming the universality of the phase transition.
       
        \begin{table}[htbp]
            \centering
             \caption{\textbf{Critical properties for the $\rm SU(3)$ Hubbard model probed under different staggered flux $\phi$.}}
            \vspace{1em}
            \begin{tabularx}{\textwidth}{l *{4}{>{\centering\arraybackslash}X}}
               \toprule
                \textbf{$\bm{\phi}$} & \textbf{$\bm{U_{c}}$} & \textbf{$\bm{\nu^{-1}}$} & \textbf{$\bm{\eta_\phi}$} & \textbf{$\bm{\eta_\psi}$} \\
                \midrule
                0.05$\pi$ & 0.347(4) & 0.67(1) & 0.51(3) & 0.16(2)\\
                0.075$\pi$ & 1.10(5) & 0.68(5) & 0.55(5) & 0.15(3) \\
                0.1$\pi$ & 2.0(3) & 0.58(7) & 0.58(10) & 0.16(3) \\
                \bottomrule
            \end{tabularx}
            \label{tab:SU3_critical}
        \end{table}

    \subsection{The new universality class}
    \label{sec:S4-D}

        In three-dimensional classical systems or 2+1 dimensional quantum systems, the most typical O($N$) universality classes widely describe a variety of important phase transitions, such as the gas-liquid transition of simple gases, the superfluid transition of liquid helium, and the Heisenberg transition of ferromagnets. Specifically, the $N=1$ case is referred to as the Ising universality class, $N=2$ as the XY universality class, and $N=3$ as the Heisenberg universality class. Their critical behaviors can be simply described by purely bosonic $\phi^4$ theories with O($N$) symmetries.  
        
        When the boson order parameter fields are coupled to Dirac fermions, the critical properties are significantly modified by the gapless fermion fluctuations, which leads to the definition of the ``chiral versions" of the above universality classes~\cite{Rosenstein1993PLB}, namely the \textit{O($N_{\text{b}}$)} universality classes. 
        To avoid confusion, we here use $N$ to denote the dimension of the classical order parameter, $N_{\text{b}}$ to denote the dimension of the bosonic order parameter, and $N_{\text{f}}$ to denote the number of fermion flavors.
        For example, quantum phase transitions with a $Z_2$ charge-density-wave order parameter belong to the \textit{chiral Ising} universality class; those with an O(2), or equivalently U(1), superconducting order parameter belong to the \textit{chiral XY} universality class; and those with an O(3) antiferromagnetic order parameter belong to the \textit{chiral Heisenberg} universality class. The corresponding low-energy effective field theories at criticality are obtained by coupling the $\phi^4$ bosonic theory of the order parameter to chiral fermion fields through Yukawa terms, and can generally be written as  
        \begin{equation}
            \mathcal{L}
            = \text{tr} \bigl(\bar{\psi}\slashed{\partial}\psi\bigr)
            + g\, \text{tr} \bigl(\bar{\psi}\Phi\psi\bigr)
            + \text{tr} \bigl((\partial_\mu \Phi)^2\bigr)
            + r\, \text{tr} \bigl(\Phi^2\bigr)
            + \lambda\, \bigl(\text{tr} (\Phi^2)\bigr)^2
            \, ,
            \label{eq:GNY}
        \end{equation}
        where $\psi$ is the fermion spinor field, and $\text{tr}$ denotes the trace over the $N_{\text{f}}$ flavors of chiral fermions with SU($N_{\text{f}}$) symmetry. The operator $\Phi=\sum_{i=1}^{N_{\text{b}}} \phi_i L_i$ spans an $N_{\text{b}}$-dimensional linear representation of the SU($N_{\text{f}}$) algebra, with $N_{\text{b}}$ linearly independent bosonic components, and the $N_{\text{f}}$-dimensional matrices $L_i$ form an orthogonal basis of this representation, such that $\text{tr}(\Phi^2)=\sum_{i=1}^{N_{\text{b}}} \phi_i^2$. This theory is the well-known Gross-Neveu-Yukawa theory. Therefore, these chiral O($N_{\text{b}}$) universality classes are collectively referred to as the \textit{Gross-Neveu universality classes}. Their order parameters correspond to the $N_{\text{b}}$-dimensional real linear representation of the fermionic SU(2) algebra, which happens to carry a faithful representation of the O($N_{\text{b}}$) group. Consequently, the bosonic fields in the theory also possess O($N_{\text{b}}$) symmetry, establishing a one-to-one correspondence with the usual O($N$) universality classes. 
        For instance, for the $N_{\text{b}}=3$ chiral Heisenberg universality class in an SU($N_{\text{f}}=2$) system, Eq.~\eqref{eq:GNY} can be written in the more familiar form  
        \begin{equation}
            \mathcal{L}
            = \bar{\psi}\slashed{\partial}\psi
            + g\, \bar{\psi}\kuohao{\vec{\phi}\cdot\vec{\sigma}}\psi
            + \abs{\partial_\mu \vec{\phi}}^2
            + r\, \abs{\vec{\phi}}^2
            + \lambda\, \abs{\vec{\phi}}^4
            \, ,
            \label{eq:chiralHeisenberg}
        \end{equation}
        where $\vec{\phi}=(\phi_1,\phi_2,\phi_3)$ is an O(3) vector and $\vec{\sigma}=(\sigma_x,\sigma_y,\sigma_z)$ are the Pauli matrices, forming an orthogonal basis for the three-dimensional linear representation of the SU(2) algebra. 
         We summarize in Table~\ref{tab:universality-classes-new} the theoretical symmetries and the vacuum manifolds (ground-state degeneracy manifolds) after spontaneous symmetry breaking corresponding to various chiral O($N_{\text{b}}$) universality classes.

        \begin{table}[bt]
            \centering
            \caption{Comparison of the symmetries of different Gross-Neveu universality classes. The symmetries of the theories and the vacuum manifolds (ground-state degeneracy manifolds) after spontaneous symmetry breaking are listed. Each chiral O($N_{\text{b}}$) universality class corresponds to an O($N$) universality class with the same symmetries and vacuum manifold. In contrast, the new chiral SU(3) universality class reported in our manuscript has no corresponding O($N$) universality class.}
            \vspace{1em}
            \setlength{\tabcolsep}{12pt}
            \begin{tabular}{lll}
                \hline
                Gross-Neveu classes & Symmetry & Vacuum manifold \\
                \hhline{===}
                \multirow{3}{*}{
                    chiral O($N_{\text{b}}$) 
                    $\left\{
                    \begin{array}{l}
                        \text{chiral Ising} \\
                        \text{chiral XY} \\
                        \text{chiral Heisenberg}
                    \end{array}
                    \right.$
                }
                & \multirow{3}{*} {O($N_{\text{b}}$) 
                    $\left\{
                    \begin{array}{l}
                        \text{$Z_2$, $N_{\text{b}}=1$ } \\
                        \text{$O(2)$, $N_{\text{b}}=2$} \\
                        \text{$O(3)$, $N_{\text{b}}=3$}
                    \end{array}
                    \right.$
                    }
                & S$^0$ \\
                & & S$^1$ \\
                & & S$^2$ \\
                \hline
                ~chiral SU(3) & $\text{SU(3)}\times \text{Z}_2$ & $\frac{\text{SU(3)}\times \text{Z}_2}{\text{SU(2)}\times \text{U(1)}}\kuohao{\simeq\mathbb{CP}^2\times \text{Z}_2}$ \\
                \hline
            \end{tabular}
            \label{tab:universality-classes-new}
        \end{table}

        The Gross-Neveu universality class describes the universal physics in which massless Dirac fermions acquire a mass through spontaneous symmetry breaking. Although the known Gross-Neveu universality classes possess an O($N$) order-parameter structure, considering the intrinsic spinor nature of fermions, the Gross-Neveu universality class should not simply be regarded as the chiral counterpart of the classical O($N$) universality class.
        A fundamental question of broad interest across statistical physics, condensed matter physics, and high-energy physics is whether fermionic Gross-Neveu criticality can arise that goes beyond the O($N$) counterparts.

\begin{figure}[bt]
            \centering
            \includegraphics{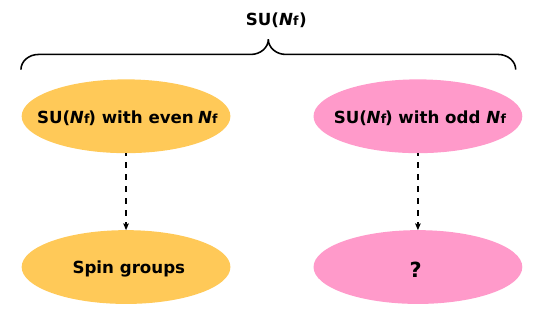}
            \caption{For odd $N_{\text{f}}$, the SU($N_{\text{f}}$) group cannot be isomorphic to a spin group, which may give rise to new Gross-Neveu universality classes that more directly manifest the spinor nature of fermions.}
            \label{fig:FIG_SUN}
        \end{figure}

        We find that SU($N_{\text{f}}$) Dirac fermion systems with with odd $N_{\text{f}}$ offer precisely such a pathway. From the perspective of the structure of symmetry group algebras, the key distinction between odd and even $N_{\text{f}}$ lies in the fact that SU($N_{\text{f}}$) groups with odd $N_{\text{f}}$ cannot be isomorphic to spin groups, which may allow for the emergence of ``more fermionic'' universality classes, as illustrated in Fig~\ref{fig:FIG_SUN}. For even $N_{\text{f}}$, however, at small values of $N_{\text{f}}$, such as SU(2) and SU(4), the groups are locally isomorphic to O(3) and O(6), respectively; while at large $N_{\text{f}}$, the system typically does not favor breaking SU($N_{\text{f}}$) symmetry to form spin order, but instead tends to form a valence-bond solid~\cite{Sachdev1989prl,Sachdev1990prb}. Indeed, all Gross-Neveu transitions studied so far for even $N_{\text{f}}$ fall back into the conventional chiral O($N_{\text{b}}$) universality classes~\cite{Qi2022prb,Wu2014prl,Wang2018prb,Meng2019prx,Meng2019prb,Xu2020prb}. Therefore, investigating the case of odd $N_{\text{f}}$ is of fundamental importance for understanding fermionic phase transitions. 

        We clearly identify the first nontrivial fermionic Gross-Neveu criticality that goes beyond the O($N$) order-parameter structures. The continuous transition from the SU(3) DSM to the $\lambda_8$-AFM phase reported in our manuscript belongs to the Gross-Neveu universality class family but does not fall into any of the known chiral Ising, chiral XY, or chiral Heisenberg classes. 
        
        In the saturated $\lambda_8$-AFM phase, two flavors of fermions (referred to as flavors 1 and 2) are localized in one sublattice (designated as the A lattice), while the remaining flavor of fermions (referred to as flavor 3) is localized in the other sublattice (designated as the B lattice), as illustrated in Fig.~\ref{fig:SU3}\subref{a} of the main text.
        It is evident that the \nota{\mathrm{Z}_2} symmetry between the AB sublattices is significantly broken.
        The diagonal \nota{\lambda_8} generator induces a global U(1) transformation, while the generators \nota{\lambda_1,\lambda_2,\lambda_3} act solely on the subspace of flavors 1 and 2, generating a closed SU(2) transformation that only mixes these two flavors.
        Since these four generators all commute with the order parameter operator, the \nota{\mathrm{SU(2)}\times\mathrm{U(1)}} is the largest symmetry group that remains invariant under the $\lambda_8$-AFM order parameter after spontaneous symmetry breaking. The other four generators of the SU(3) group, \nota{\lambda_4,\lambda_5,\lambda_6,\lambda_7}, produce transformations that alter the direction of the AFM order parameter $\left\{m_n\right\}$ in a compact manifold, leading to other $\lambda_8$-AFM degenerate ground states, which are four independent Goldstone modes. All degenerate ground states span a 4-dimensional ground state degeneracy space, and these states can be mapped one-to-one onto points on the manifold $\frac{\text{SU(3)}\times \text{Z}_2}{\text{SU(2)}\times \text{U(1)}}$. 
        Under SU(3) transformations, this order parameter transforms according to the 8-dimensional adjoint representation of SU(3) and spans the 4-dimensional $\frac{\text{SU(3)}\times \text{Z}_2}{\text{SU(2)}\times \text{U(1)}}$ manifold, which is nonlinearly embedded in $\mathbb{R}^8$. SU(3) cannot be isomorphic or locally isomorphic to any other classical linear groups, which is in sharp contrast to the cases of even $N_{\text{f}}$ such as SU(2) and SU(4). Therefore, the new SU(3) antiferromagnetic order parameter we have discovered does not have a corresponding O($N_{\text{b}}$) order parameter. We temporarily refer to this new universality class as the \textit{chiral SU(3)} universality class. As shown in the comparison in Table~\ref{tab:universality-classes-new} summarizes the differences of this new universality class from the previously known ones in terms of symmetry and vacuum manifold. 
        In addition to the symmetry analysis, our QMC results also provide an unbiased numerical determination of the critical exponents of this new universality class. In Table~\ref{tab:SU3_comparison}, we present a comparison of its critical exponents with those of the chiral Ising, chiral XY, and chiral Heisenberg universality classes, which more conclusively rules out the possibility that this new universality class belongs to those existing universality classes.

        \begin{table}[bt]
            \centering
             \caption{\textbf{Comparison of critical exponents for different Gross-Neveu-Yukawa universality classes in $d=2+1$ with $N_\mathrm{f}=6$.} The first row, chiral SU(3) denotes the universality class for SU(3) Dirac fermion Hubbard model, with exponents determined from nonequilibrium PQMC.}
            \vspace{1em}
            \begin{tabularx}{\textwidth}{l *{3}{>{\centering\arraybackslash}X}}
               \toprule
                \textbf{Universality class} & \textbf{$\bm{\nu^{-1}}$} & \textbf{$\bm{\eta_\phi}$} & \textbf{$\bm{\eta_\psi}$} \\
                \midrule
                chiral SU(3) (this work) & 0.68(5) & 0.55(5) & 0.15(3) \\
                chiral Heisenberg ($4-\epsilon$, 2nd order)\cite{Rosenstein1993PLB} & 1.478 & 1.023 & 0.058 \\
                chiral XY ($4-\epsilon$, 2nd order)\cite{Rosenstein1993PLB} & 1.809 & 0.698 & 0.082 \\
                chiral Ising ($4-\epsilon$, 2nd order)\cite{Rosenstein1993PLB} & 0.750 & 0.865 & 0.011 \\
                chiral Ising (FRG)\cite{Herbut2014prb} & 0.993 & 0.912 & 0.013 \\
                \bottomrule
            \end{tabularx}
            \label{tab:SU3_comparison}
        \end{table}

        \begin{figure}[bt]
            \centering
            \includegraphics{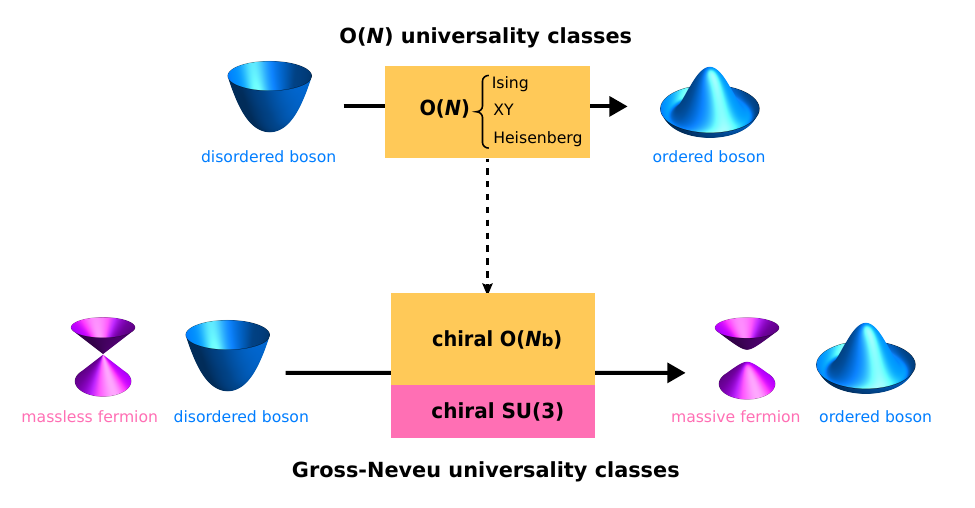}
            \caption{Schematic illustration of the relationship between O($N$) universality classes and Gross-Neveu universality classes. The O($N$) universality classes, including Ising, Heisenberg, and XY, describe spontaneous symmetry breaking with an O($N$) order parameter. The Gross-Neveu universality classes describe the situation where, along with spontaneous symmetry breaking of the bosonic order parameter, massless Dirac fermions also spontaneously acquire a mass. 
            Within the Gross-Neveu universality classes, the traditional chiral O($N$) universality classes correspond one-to-one to the O($N$) universality classes, whereas the chiral SU(3) universality class in the Gross-Neveu universality classes does not possess an O($N$) order-parameter structure and instead represents a fermion-intrinsic universality class.}
            \label{fig:FIG_GNY}
        \end{figure}

        The nontrivial new Gross-Neveu universality class we discovered represents the first minimal example of a Gross-Neveu transition without a classical O($N$) correspondence. As illustrated in Fig.~\ref{fig:FIG_GNY}, our results demonstrate that the Gross-Neveu universality class encompasses a broader and nontrivial set with intrinsic fermionic characteristics.
        It is highly worthwhile in the future to further investigate the low-energy effective field theory of this phase transition. Since the bosonic field does not possess O($N_{\text{b}}$) symmetry, its form is likely different from the previous expression in Eq.~\eqref{eq:GNY}. We conjecture that it may take the following form:
        \begin{equation}
            \mathcal{L}
            = \text{tr} \kuohao{\bar{\psi}\slashed{\partial}\psi}
            + g\, \text{tr} \kuohao{\bar{\psi}\Phi\psi}
            + \text{tr} \fkuohao{\kuohao{\partial_\mu \Phi}^2}
            + r\, \text{tr} \kuohao{\Phi^2}
            + \lambda\,  \fkuohao{\text{tr} \kuohao{\Phi^2}}^2
            + a\, \fkuohao{\text{tr} \kuohao{\Phi^3}}^2
            + c\, \fkuohao{\text{tr} \kuohao{\Phi^2}}^4
            \, ,
            \label{eq:GNY-SU3}
        \end{equation}
        where $\Phi=\sum_{i=1}^{N_{\text{b}}} \phi_i\, \frac{\lambda_i}{2}, N_{\text{b}}=8$ spans the 8-dimensional adjoint representation of SU(3). Compared with Eq.~\eqref{eq:GNY}, the marginal term $a\, \fkuohao{\text{tr} \kuohao{\Phi^3}}^2$ is introduced to reduce the SO(8) symmetry of the bosonic field to $\text{SU(3)}\times \text{Z}_2$ symmetry, while the irrelevant term $c\, \fkuohao{\text{tr} \kuohao{\Phi^2}}^4$ with $c>0$ is also introduced to ensure vacuum stability. Due to the SU(3) algebra identity $\text{tr}(\Phi^3)=3\det \Phi$, only the $\phi_8$ component of the order parameter along the $\lambda_8$ direction with full rank contributes a nonzero value to $\fkuohao{\text{tr} \kuohao{\Phi^3}}^2$. Therefore, when $a<0$, the SU(3)-symmetric bosonic field tends to spontaneously break toward the $\lambda_8$ direction. 
        This is very different from the O($N$) symmetry breaking triggered by $r<0$ in the $\phi^4$ theory.
        Whether such a Gross-Neveu-Yukawa theory can capture the new universality class we discovered (for example, yielding critical exponents consistent with our numerical results) remains an open question for future study.
        Additionally, such an order parameter that carries an SU(3) bilinear representation rather than an O($N_{\text{b}}$) linear representation may also be constructed through a recent theoretical framework with tensorial criticality~\cite{Herbut2024prb,Ray2024prb,Herbut2025prb}, which is also worth exploring.

    \subsection{Sign problem behaviors and computational efficiency}

        \begin{figure}[htbp]
            \centering
            \includegraphics{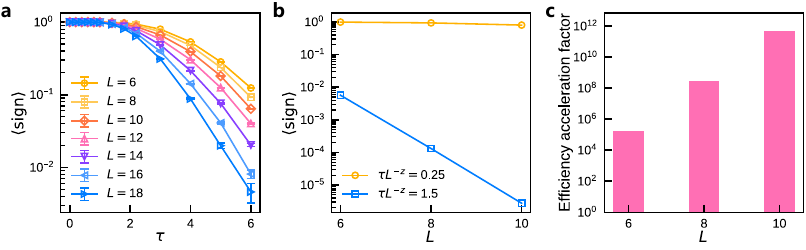}
            \caption{
                \textbf{The sign problem in SU(3) Hubbard model at the QCP $\phi=0.1,U=2.0(3)$.} 
                \textbf{a}, The average sign with different $L$ and $\tau$ in short-time stage. 
                \textbf{b}, Comparison for the average sign between short-time stage ($\tau=0.25L^z$) and equilibrium stage ($\tau=1.5L^z$).
                \textbf{c}, Efficiency gain of QMC with the short-time method ($\tau=0.25L^z$) compared to equilibrium QMC ($\tau=1.5L^z$) for different sizes $L$.
            }
            \label{fig:SU3_efficiency}
        \end{figure}

        For this model, we set the parameters at the critical point $U=2.0(3)$, $\phi=0.1\pi$. The average sign as a function of imaginary time \nota{\tau} and size $L$ is shown in Fig.~\ref{fig:SU3_efficiency}\subref{a}. The sign problem for this model is significantly more severe than for the previous two models. For \nota{\tau =0.25 L^z}, the average sign is approximately \nota{10^{-1}\sim10^{-2}}, meaning that compared to the sign-problem-free case, $10$ to $100$ times more computational resources are needed to obtain reliable results. From Figs.~\ref{fig:SU3_relaxation_005} and \ref{fig:SU3_relaxation_01} and Fig.~\ref{fig:SU3} in the main text, it can be seen that \nota{\tau =0.25 L^z} is in the nonequilibrium scaling region controlled by the critical point. Even though the ground state is not reached, its nonequilibrium scaling still reflects the quantum criticality of the ground state. In equilibrium QMC studies, it is typically necessary to set the imaginary time \nota{\tau} to more than $1.5$ times the size \nota{L^z} to reach the ground state. However, evolving for such a long time, the average sign decays to approximately \nota{10^{-5}\sim10^{-6}}. In Fig.~\ref{fig:SU3_efficiency}\subref{b}, we compare the average sign for \nota{\tau =0.25 L^z} and \nota{\tau =1.5 L^z}. Since the computational error (according to Eq.~\ref{eq:sign_err}) is inversely proportional to the average sign $\braket{\mathrm{sign}}$ (according to Eq.~\ref{eq:sign_err}), we can measure the difference in computational efficiency by multiplying the ratio of the average sign $\braket{\mathrm{sign}}$ by the length of imaginary-time evolution, as follows:
        \begin{equation}
            \mathrm{Efficiency~acceleration~factor} = \left(\frac{{1/\braket{\mathrm{sign}}_\mathrm{eq.}}}{{1/\braket{\mathrm{sign}}_\mathrm{neq.}}} \right)^2 \frac{\tau_\mathrm{eq.}}{\tau_\mathrm{neq.}},
        \end{equation}
        where the nonequilibrium evolution time is $\tau_\mathrm{neq.}=0.25L^z$, and the equilibrium evolution time is $\tau_\mathrm{eq.}=1.5L^z$. $\braket{\mathrm{sign}}_\mathrm{neq.}$ and $\braket{\mathrm{sign}}_\mathrm{eq.}$ are their corresponding average signs, as shown in Fig.~\ref{fig:SU3_efficiency}\subref{b}. 
        We specifically compared the differences in computational efficiency for several system sizes, as shown in Fig.~\ref{fig:SU3_efficiency}\subref{c}. The computational resources required by the nonequilibrium method are only a few millionths of those required by the equilibrium method, and this efficiency improvement roughly grows exponentially with the system size. Consequently, our nonequilibrium method enables the QMC simulation on SU(3) Hubbard model with large system size, which is not accessible in previous unbiased numerical approaches.  

\end{document}